\documentclass[10pt,twocolumn,letterpaper]{article}

\usepackage{cvpr}              

\usepackage{tcolorbox}

\usepackage{placeins}
\usepackage{float} 

\definecolor{cvprblue}{rgb}{0.21,0.49,0.74}
\usepackage[pagebackref,breaklinks,colorlinks,allcolors=cvprblue]{hyperref}
\usepackage{colortbl}
\usepackage{placeins}
\def\paperID{*****} 
\def\confName{CVPR}
\def\confYear{2026}

\usepackage{placeins}

\AtBeginDocument{%
    \AddToShipoutPicture*{
        \AtPageUpperLeft{%
            \put(\LenToUnit{0.5\paperwidth-3cm},\LenToUnit{-2.5cm}){%
                \includegraphics[height=1.3cm]{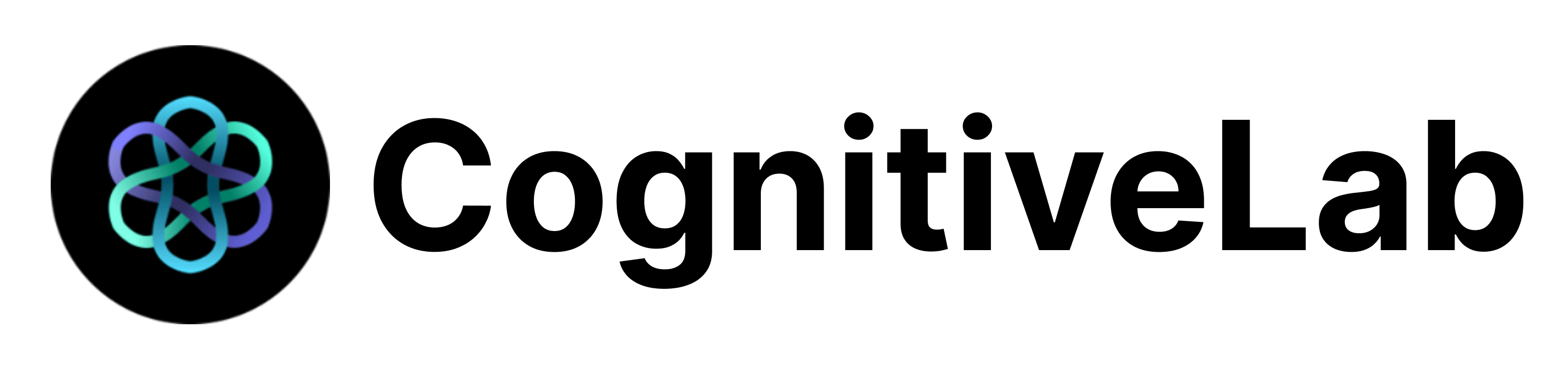}%
            }%
        }%
    }%
}
\title{M3DR: Towards Universal Multilingual Multimodal Document Retrieval}

\author{
Adithya S Kolavi\\
CognitiveLab\\
Bengaluru, India\\
{\tt\small adithyaskolavi@gmail.com}
\and
Vyoman Jain\\
CognitiveLab\\
Bengaluru, India\\
{\tt\small vyomanjain@gmail.com}
}
\begin{document}
\maketitle
\begin{abstract}
Multimodal document retrieval systems have shown strong progress in aligning visual and textual content for semantic search. However, most existing approaches remain heavily English-centric, limiting their effectiveness in multilingual contexts. In this work, we present M3DR (Multilingual Multimodal Document Retrieval), a framework designed to bridge this gap across languages, enabling applicability across diverse linguistic and cultural contexts. M3DR leverages synthetic multilingual document data and generalizes across different vision-language architectures and model sizes, enabling robust cross-lingual and cross-modal alignment. Using contrastive training, our models learn unified representations for text and document images that transfer effectively across languages. We validate this capability on 22 typologically diverse languages, demonstrating consistent performance and adaptability across linguistic and script variations. We further introduce a comprehensive benchmark that captures real-world multilingual scenarios, evaluating models under monolingual, multilingual, and mixed-language settings. M3DR generalizes across both single dense vector and ColBERT-style token-level multi-vector retrieval paradigms. Our models, NetraEmbed and ColNetraEmbed achieve state-of-the-art performance with $\sim$150\% relative improvements on cross-lingual retrieval.
\end{abstract}
\section{Introduction}
\label{sec:intro}

The exponential growth of digital documents across global enterprises, research institutions, and digital libraries has created an urgent need for effective multilingual document retrieval systems. While text-based retrieval has achieved remarkable success across languages ~\cite{dense_passage_retrieval, contriever}, retrieval based on traditional OCR-based pipelines face significant challenges: information loss from discarding visual elements (charts, diagrams, layout), brittleness with diverse fonts and scripts, and cascading errors particularly severe in low-resource languages.

Recent vision-based approaches like ColPali~\cite{colpali2024} have demonstrated promising results by directly encoding document images using vision language models (VLMs), thereby eliminating OCR dependencies and preserving rich visual–textual information. However, these systems remain predominantly English-centric, and our preliminary analysis shows that they perform very poorly on multilingual content, a critical limitation in a world where documents span hundreds of languages.

This work addresses a fundamental research question: \textit{Can we develop universal document retrievers that maintain high performance across typologically diverse languages without sacrificing English competitiveness?}

\begin{figure}[!h]
\centering
\includegraphics[width=\columnwidth]{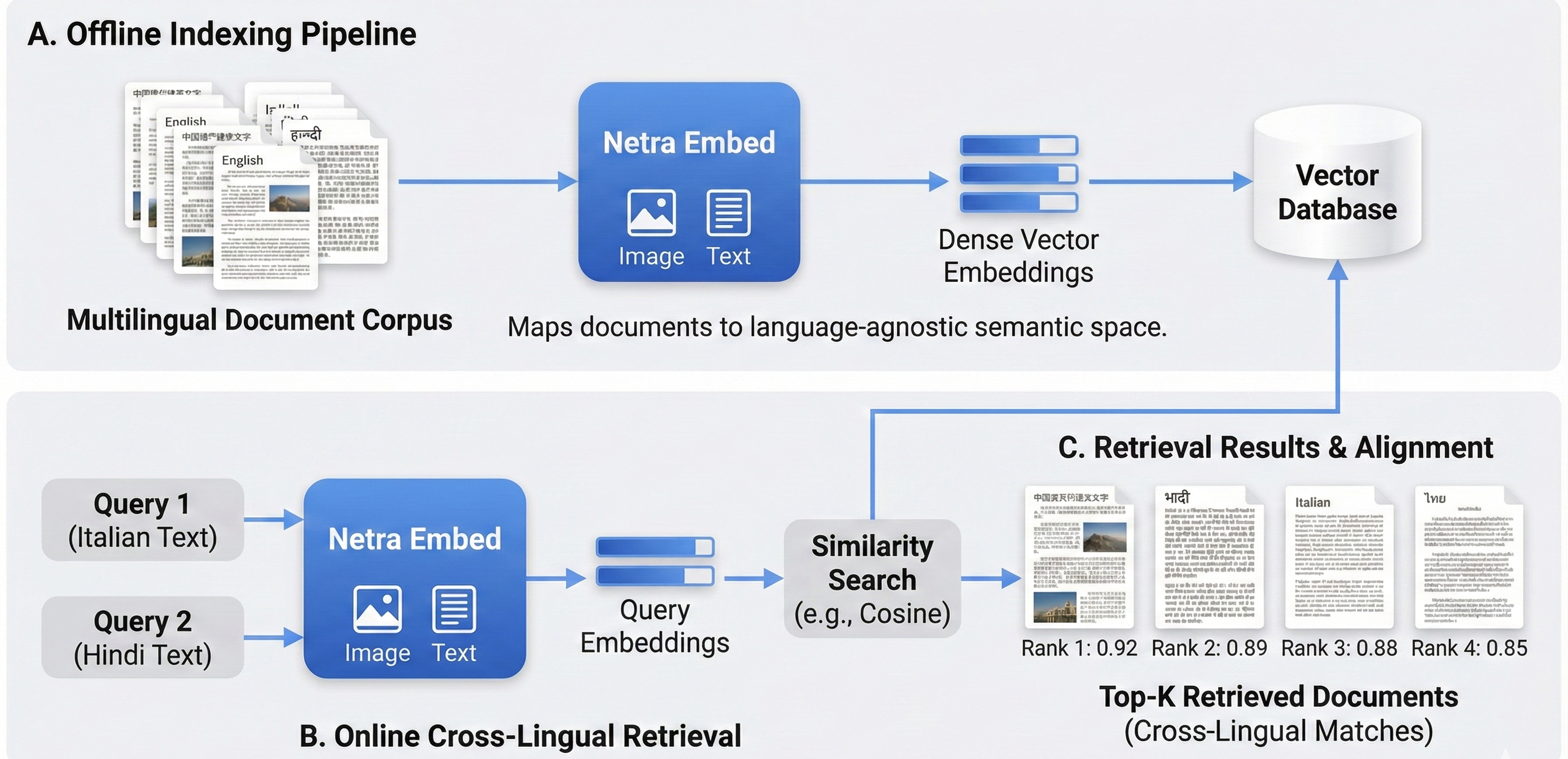}
\caption{Overview of NetraEmbed, our multilingual multimodal document embedding model. (A) Offline indexing encodes documents into dense vectors in a shared semantic space, (B) online retrieval processes cross-lingual queries, and (C) results show effective matching across diverse scripts and languages.}
\label{fig:netraembed_flow}
\end{figure}

We present \textbf{M3DR}, a scalable framework for training multilingual multimodal document retrievers across 22 typologically diverse languages spanning Latin, Devanagari, Dravidian, CJK, and other script families. M3DR generalizes across both single dense vector and ColBERT-style multi-vector retrieval paradigms. With our single dense vector model, we achieve 152\% relative improvement over baselines on cross-lingual retrieval (0.716 vs 0.284 NDCG@5) establishing new state-of-the-art performance for multilingual multimodal document retrieval.

\noindent We make \textbf{three key contributions}:
\begin{itemize}
\item We develop a synthetic data generation pipeline for generating multilingual multimodal document retrieval data at scale. Our approach combines layout-aware document translation, authentic typography rendering with language-specific fonts, and VLM-based query synthesis (Llama 3.1 90B Vision~\cite{llama3}, Llama 4 Scout~\cite{llama4}), producing $\sim$1M parallel document images across 22 languages

\item We introduce a ocomprehensive multilingual multimodal document retrieval benchmark \textbf{Nayana IR Bench}\footnote{\href{https://huggingface.co/collections/Cognitive-Lab/nayanair-bench}{NayanaIR bench}} with 23 datasets (1 cross lingual plus 22 monolingual), $\sim$28K document images, and $\sim$5.4K queries in BEIR compatible format~\cite{beir2021}. This benchmark supports standardized evaluation across diverse script families for both cross lingual and monolingual retrieval capabilities.

\item We release two 4B parameter models demonstrating M3DR's generalization across retrieval paradigms. \\
\textbf{NetraEmbed}\footnote{\href{https://huggingface.co/Cognitive-Lab/NetraEmbed}{NetraEmbed}} (Netra is Sanskrit for eye or vision), a single dense vector model using Matryoshka representation learning that supports multiple output dimensions (768, 1536, 2560), and \textbf{ColNetraEmbed}\footnote{\href{https://huggingface.co/Cognitive-Lab/ColNetraEmbed}{ColNetraEmbed}}, a ColBERT style multi vector variant. NetraEmbed reaches state of the art results on cross lingual and mono lingual retrieval while keeping strong English performance.
        

\end{itemize}

\section{Related Work}
\label{sec:related}

Our work draws upon and extends several research threads. We organize related work into five key areas: visual document understanding, vision-based document retrieval, multimodal retrieval systems, multilingual embeddings, and training strategies.

\noindent\textbf{Visual Document Understanding.} Traditional document understanding relies on OCR followed by text processing~\cite{layoutlmv3, nougat2023}, suffering from information loss and error propagation. Recent vision-language models enable end-to-end understanding. mPLUG-DocOwl~\cite{mplug_docowl2024} introduced structure-aware learning for OCR-free document understanding across documents, tables, and charts. mPLUG-DocOwl2~\cite{mplug_docowl2_2024} proposed high-resolution compression, reducing visual tokens from thousands to 324 while maintaining performance. HRVDA~\cite{hrvda2024} addressed high-resolution challenges through content and instruction filtering. For long documents, LongDocURL~\cite{longdocurl2024} introduced comprehensive benchmarks spanning 33K+ pages, while MMLongBench-Doc~\cite{mmlongbench2024} revealed performance gaps in current VLMs on lengthy PDFs (avg. 49.4 pages). SlideVQA~\cite{slidevqa2023} addressed multi-image document VQA with slide decks. While these works demonstrate impressive capabilities, they primarily focus on question-answering rather than retrieval, and evaluations center on English. M3DR complements this by enabling multilingual document retrieval as a critical first step for document-centric RAG.

\noindent\textbf{Vision-Based Document Retrieval.} ColPali~\cite{colpali2024} pioneered vision-based document retrieval by adapting late-interaction mechanisms from ColBERT~\cite{colbert2020} to visual documents, achieving substantial improvements over OCR-based pipelines on the ViDoRe~\cite{macé2025vidorebenchmarkv2raising} benchmark. A reproducibility study~\cite{colpali_repro2025} confirmed ColPali's effectiveness and provided insights into query-patch matching. ModernVBERT~\cite{modernvbert2025} proposed a compact 250M-parameter encoder optimized for document retrieval, competitive with 10× larger models through principled design. Document Screenshot Embedding~\cite{dse2024} explored using screenshots as unified input, demonstrating versatility across document types. Guided Query Refinement~\cite{gqr2025} proposed test-time optimization for query embeddings, while EDJE~\cite{edje2025} introduced efficient discriminative joint encoders for large-scale reranking. While these methods demonstrate strong English performance, none systematically address multilingual scenarios. M3DR extends vision-based retrieval to 22 languages, validating that these approaches generalize across linguistic and script boundaries.

\noindent\textbf{Multimodal Retrieval and RAG.} Universal multimodal retrieval aims to handle diverse modalities. UniIR~\cite{uniir2023} introduced instruction-guided retrieval with the M-BEIR benchmark. MM-Embed~\cite{mm_embed2024} trained multimodal embedders using MLLMs with modality-aware hard negative mining. GME~\cite{gme2024} improved upon MM-Embed through high-quality fused-modal data synthesis. U-MARVEL~\cite{umarvel2025} provided comprehensive study of key factors including progressive transition and hard negative mining. LamRA~\cite{lamra2024} re-purposed generative LMMs for retrieval through unified structure learning. For document-specific RAG, M3DocRAG~\cite{m3docrag} combined ColPali with MLMs for multi-page, multi-document QA. VisRAG~\cite{visrag2024} established vision-based RAG pipelines, demonstrating 20-40\% gains over text-based RAG. VISA~\cite{visa2024} enhanced RAG with visual source attribution using bounding boxes. Benchmarking efforts like MMDocIR~\cite{mmdocir2025}, REAL-MM-RAG~\cite{real_mm_rag2025}, and UDA~\cite{uda2024} have been crucial. M3DR fills the gap by combining document-specific understanding with comprehensive multilingual support.

\noindent\textbf{Multilingual and Cross-Lingual Retrieval.} Text-based multilingual embeddings like mBERT~\cite{mbert2019}, XLM-R~\cite{xlm_roberta}, and LaBSE~\cite{labse2022} achieve cross-lingual transfer but discard visual information. The GTE series~\cite{gte2023} employed multi-stage contrastive learning on large-scale multilingual data. Jina-embeddings-v3~\cite{jina_v3_2024} introduced task-specific LoRA adapters, while Jina-v4~\cite{jina_v4_2025} extended to multimodal multilingual retrieval with separate text and image encoders, achieving strong performance on JVDR benchmark. However, separate encoders may miss fine-grained visual-textual interactions that vision-based approaches capture. VLM2Vec~\cite{vlm2vec2024} proposed converting VLMs into embedding models through contrastive training on MMEB. xVLM2Vec~\cite{xvlm2vec2025} extended to multilingual scenarios through self-knowledge distillation, though limited to European languages. BGE-M3~\cite{bge_m3_2024} supports 100+ languages over text. Unlike these works which focus on text-only multilingual retrieval or limited-language multimodal scenarios, M3DR systematically addresses multilingual visual document retrieval across diverse scripts.

\noindent\textbf{Training Strategies.} Modern dense retrievers rely on contrastive learning with InfoNCE loss~\cite{infonce2018} and hard negative mining. NV-Retriever~\cite{nv_retriever2024} introduced positive-aware mining for effective false negative removal. LLaRA~\cite{llara2023} adapted LLMs for retrieval through embedding-based auto-encoding. Llama2Vec~\cite{llama2vec2024} demonstrated unsupervised adaptation of Llama-2 for dense retrieval. DRAMA~\cite{drama2025} leveraged LLMs for diverse data augmentation. Matryoshka representation learning~\cite{matryoshka2022} enables flexible embedding dimensions, critical for deployment efficiency. M3DR builds upon these strategies, conducting systematic ablations on loss functions, negative sampling, and Matryoshka training, identifying optimal configurations for multilingual multimodal document retrieval.

\section{Training Data and Benchmark}
\label{sec:dataset}

A critical challenge in developing multilingual multimodal document retrieval systems is the scarcity of high-quality training data and comprehensive evaluation benchmarks. In this section, we describe our approach to addressing this challenge through: (1) a scalable synthetic data generation pipeline that creates parallel multilingual document corpora for training, (2) query synthesis using large vision-language models, and (3) construction of the \textbf{Nayana-IR Benchmark} for standardized evaluation.

\subsection{Training Dataset Construction}

\subsubsection{Synthetic Parallel Corpus Generation}

Building on the Nayana framework~\cite{kolavi-etal-2025-nayana, kolavi2025nayana_iccv}, which demonstrated the effectiveness of layout-aware synthetic data generation for adapting vision-language models to low-resource languages, we extend this approach to create a large-scale multilingual parallel corpus for document retrieval. While the original Nayana work focused on OCR and document understanding tasks, we adapt their synthesis pipeline for retrieval by generating document images paired with diverse queries across 22 languages.

\noindent\textbf{Source Documents.} We curate ~50,000 diverse English documents images spanning scientific papers, technical reports, educational materials, business documents, and forms from publicly available sources.

\begin{figure*}[t]
  \centering
  \includegraphics[width=\linewidth]{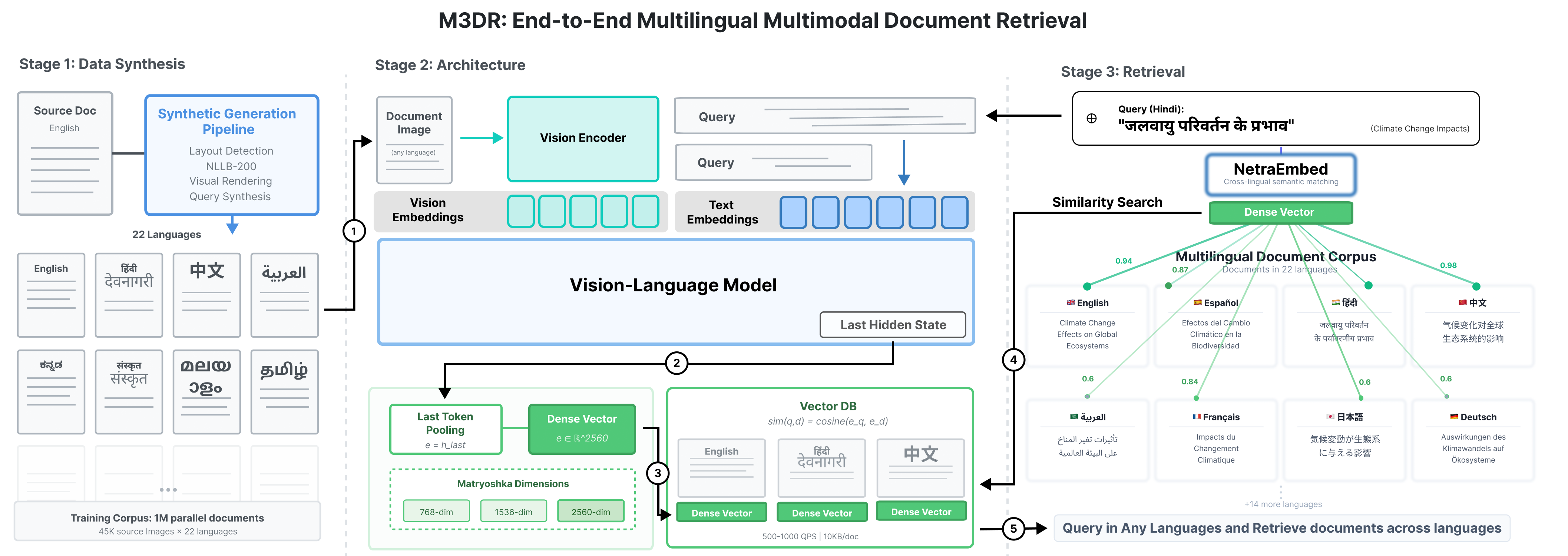}
  \caption{\textbf{M3DR Framework Overview.} Our complete pipeline encompasses synthetic data generation (layout detection, neural translation to 22 languages, visual rendering with authentic typography), query synthesis using large VLMs, dense embedding model training with Matryoshka representation learning, and multilingual document retrieval across diverse script families.}
  \label{fig:framework_overview}
\end{figure*}

\noindent\textbf{Layout-Aware Translation.} Following the Nayana approach, our pipeline (Figure~\ref{fig:framework_overview}) employs: (1) \textit{Layout Detection}: DocLayout-YOLO~\cite{doclayout_yolo}, Docling~\cite{docling2024} extract text regions, visual elements, tables, and layout metadata. (2) \textit{Neural Translation}: Context-aware translation using NLLB-200~\cite{flores200} and language-specific models~\cite{gumma-etal-2025-towards} preserves document semantics across 22 target languages. (3) \textit{Visual Rendering}: Authentic typography with Noto Sans fonts~\cite{noto_fonts} for universal script coverage, language-specific layout rules (character spacing, line breaking, text direction), and high-resolution rendering (1024-2048px height) maintaining visual elements from source documents.

\subsubsection{Query Synthesis}

\noindent To support diverse retrieval scenarios, we generate multiple types of queries for each document image using state-of-the-art large vision-language models (Llama 3.1 90B Vision~\cite{llama3} and Llama 4 Scout~\cite{llama4}). We generate 5 diverse query types per document: (1) Basic factual questions (2×), (2) Long-answer questions (1×), (3) Multiple choice/short answer (1×), (4) Cross-paragraph reasoning (1×). LLM-based filtering ensures quality. This yields a training corpus of ~1M document images with corresponding queries across 22 languages.
\noindent\textbf{Query Distribution.} Training queries: 70\% questions, 15\% answers (reverse lookup), 15\% text snippets (keyword search).

\subsection{Nayana-IR Benchmark}

\noindent While the synthetic parallel corpus enables model training, comprehensive evaluation requires carefully curated test sets that are completely separate from training data. We introduce \textbf{Nayana-IR}, the first comprehensive benchmark for multilingual multimodal document retrieval, comprising 23 datasets across 22 languages with both cross-lingual and monolingual evaluation protocols.

\subsubsection{Benchmark Structure}

\noindent\textbf{Cross-Lingual Dataset.} 5,870 parallel document images across all 22 languages with 1,000 queries distributed uniformly ($\sim$45 per language). Evaluates: \textit{Can models retrieve documents in any language given queries in any language?} Binary relevance (score 2: exact match, score 0: non-relevant).

\noindent\textbf{Monolingual Datasets (22).} Per-language datasets with $\sim$1,000 documents and $\sim$200 queries each. Evaluates: \textit{Can models retrieve documents written in the same language given queries in that language?} Graded relevance (score 2: exact match, score 1: same document partial match, score 0: non-relevant).
\noindent\textbf{Format and Metrics.} BEIR compatible structure \cite{beir2021} enables standardized evaluation. Metrics: NDCG@5 and NDCG@10, Recall@5 and Recall@10, MAP@10, and MRR@10. The benchmark provides comprehensive coverage across script families with balanced dataset sizes: Latin script (English, Spanish, French, German, Italian), Devanagari (Hindi, Marathi, Sanskrit), Dravidian (Kannada, Telugu, Tamil, Malayalam), CJK (Chinese, Japanese, Korean), and other scripts (Arabic, Bengali, Gujarati, Odia, Punjabi, Russian, Thai). Each monolingual dataset contains approximately 1000 documents and 200 queries, while the cross lingual dataset spans all 22 languages with 5870 parallel documents and 1000 queries. In total this benchmark comprises \textbf{23 datasets, 27870 images, and 5400 queries}.

\section{Methodology}
\label{sec:method}

 Our design prioritizes: (1) \textit{architectural flexibility} to accommodate VLM backbones from 256M to 4B parameters, (2) \textit{efficiency and scalability} through single dense vector embeddings with flexible dimensionality and ColBERT-Style multi-vector model, (3) \textit{multilingual generalization} across 22 languages with diverse scripts.

\subsection{Model Architecture}
\label{sec:architectures}

\subsubsection{Single Dense Vector Model}
\label{sec:dense_models}

Single dense vector models produce fixed-size vectors enabling efficient retrieval via approximate nearest neighbor search~\cite{faiss2017, hnsw2016}. Given text query $q$ and document image $d$, we process each through a VLM backbone to obtain token-level hidden states $\mathbf{H} \in \mathbb{R}^{n \times h}$. We primarily use Gemma 3 4B-it~\cite{gemma3}, which directly outputs 2560-dimensional representations. We apply last token pooling to obtain document-level embeddings.
Embeddings are L2-normalized and similarity computed via:
\begin{equation}
\text{sim}(q,d) = \frac{\mathbf{e}_q \cdot \mathbf{e}_d}{\|\mathbf{e}_q\|_2 \|\mathbf{e}_d\|_2}
\end{equation}
We also evaluate with alternative backbones (Qwen2-VL 2B~\cite{qwen2vl2024}, SmolVLM ~\cite{marafioti2025smolvlm}), employing the same approach with their native output dimensions.

\noindent\textbf{Matryoshka Representation Learning.} To enable flexible accuracy-efficiency trade-offs, we incorporate Matryoshka learning~\cite{matryoshka2022}, training embeddings to be truncatable at dimensions 768, 1536, and 2560. During training, for each embedding $\mathbf{e} \in \mathbb{R}^{2560}$, we compute losses at three granularities:
\begin{equation}
\mathcal{L}_{\text{Matryoshka}} = \sum_{d \in \{768, 1536, 2560\}} w_d \cdot \mathcal{L}_{\text{base}}(\text{L2-norm}(\mathbf{e}[:d]))
\end{equation}
where $w_d = 1/3$. This enables post-deployment dimension selection: 768-dim (70\% storage reduction), 1536-dim (40\% reduction), or 2560-dim (maximum accuracy), requiring only embedding truncation without retraining.

\subsubsection{ColBERT-Style Multi-Vector Model}
\label{sec:late_interaction_models}

To demonstrate M3DR's generalizability across retrieval paradigms, we also develop a ColBERT-style multi-vector variant following ColPali~\cite{colpali2024}. Unlike single dense vector models that pool token representations into a single vector, ColBERT-style multi-vector models retain per-token embeddings to enable fine-grained matching.

Given query $q$ and document $d$, we process each through the VLM backbone (Gemma 3 4B-it~\cite{gemma3}) to obtain token-level representations $\mathbf{Q} \in \mathbb{R}^{n_q \times h}$ and $\mathbf{D} \in \mathbb{R}^{n_d \times h}$. For Gemma 3 4B-it, each document image produces 256 visual tokens while queries vary based on query length. Similarity is computed via late interaction using MaxSim:

\begin{equation}
\text{sim}_{\text{late}}(q,d) = \sum_{i=1}^{n_q} \max_{j=1}^{n_d} \text{cos}(\mathbf{q}_i, \mathbf{d}_j)
\end{equation}

where $\mathbf{q}_i$ and $\mathbf{d}_j$ are L2-normalized token embeddings. This enables each query token to match its most similar document token, capturing fine-grained semantic correspondences. Each document image produces 256$\times$128-dimensional embeddings compared to a single 2560-dimensional embedding for single dense vector retrieval.

\subsection{Training Objectives}
\label{sec:objectives}

\subsubsection{Loss Functions for Single Dense Vector Models}

\noindent\textbf{BiEncoderLoss (InfoNCE).} Our baseline employs InfoNCE~\cite{infonce2018} with in-batch negatives. Given batch $B$ of query-document pairs $\{(q_i, d_i^+)\}_{i=1}^B$:
\begin{equation}
\mathcal{L}_{\text{BiEncoder}} = -\frac{1}{B} \sum_{i=1}^{B} \log \frac{\exp(s_{ii}/\tau)}{\sum_{j=1}^{B} \exp(s_{ij}/\tau)}
\end{equation}
where $s_{ij} = \text{sim}(q_i, d_j)$ and $\tau=0.02$. This ranks positive $d_i^+$ higher than in-batch documents $\{d_j\}_{j \neq i}$.

\noindent\textbf{BiNegativeCELoss (Hybrid Loss).} When explicit hard negatives $\{d_i^{-,k}\}_{k=1}^K$ are available, we combine pairwise ranking with InfoNCE:
\begin{equation}
\mathcal{L}_{\text{BiNegCE}} = (1-\lambda) \cdot \mathcal{L}_{\text{pairwise}} + \lambda \cdot \mathcal{L}_{\text{InfoNCE}}
\end{equation}
\begin{equation}
\mathcal{L}_{\text{pairwise}} = \frac{1}{BK} \sum_{i=1}^{B} \sum_{k=1}^{K} \text{softplus}\left(\frac{s(q_i, d_i^{-,k}) - s(q_i, d_i^+)}{\tau}\right)
\end{equation}
where $\lambda=0.5$ balances discrimination against hard negatives with in-batch diversity, essential for cross-lingual retrieval.

\noindent\textbf{MatryoshkaBiEncoderLoss.} We wrap base losses with Matryoshka mechanism:
\begin{equation}
\mathcal{L}_{\text{Matryoshka}} = \frac{1}{3} \sum_{d \in \{768, 1536, 2560\}} \mathcal{L}_{\text{base}}\left(\text{L2-norm}(\mathbf{E}[:,:d])\right)
\end{equation}

\subsubsection{Loss Functions for ColBERT-Style Multi-Vector Models}

For the ColBERT-style multi-vector variant, we employ InfoNCE loss with in-batch negatives following ColPali~\cite{colpali2024}. Given batch $B$ of query-document pairs, the similarity is first computed via MaxSim (Equation 3), optionally normalized by query length:

\begin{equation}
\text{sim}_{\text{norm}}(q_i, d_j) = \frac{\text{sim}_{\text{late}}(q_i, d_j)}{\text{len}(q_i)}
\end{equation}

The InfoNCE loss with temperature scaling is then applied:

\begin{equation}
\mathcal{L}_{\text{late}} = -\frac{1}{B} \sum_{i=1}^{B} \log \frac{\exp(\text{sim}_{\text{norm}}(q_i, d_i^+)/\tau)}{\sum_{j=1}^{B} \exp(\text{sim}_{\text{norm}}(q_i, d_j)/\tau)}
\end{equation}

where $\tau=0.02$ is the temperature parameter. We found that the standard ColBERT loss (without explicit hard negatives) worked best for our multilingual setting, using in-batch negatives for contrastive learning. This encourages query tokens to find strong matches with relevant document regions while maintaining discrimination against other documents in the batch.

\subsection{Training Strategy}
\label{sec:training_strategy}

We investigate three progressively sophisticated strategies for our single dense vector models:

\noindent\textbf{Strategy 1: Positive-Only.} Uses only in-batch negatives $\{d_j\}_{j \neq i}$ with BiEncoderLoss. Computationally efficient baseline but most negatives are trivially distinguishable.

\noindent\textbf{Strategy 2: Document-Level Negatives.} Samples $K=3$ hard negatives from nearby pages ($p \pm \{1,2,3\}$) within the same document. These share thematic content and layout but lack specific query answers. Trained with BiNegativeCELoss ($\lambda=0.3$), encouraging fine-grained content discrimination.

\noindent\textbf{Strategy 3: Mined Hard Negatives.} Mines corpus-wide negatives via: (1) \textit{Text similarity} using BM25~\cite{bm25} and BGE-M3~\cite{bge_m3_2024} on OCR text, (2) \textit{Visual similarity} using CLIP~\cite{clip2021} and Jina-CLIP~\cite{jina_clip2024} on images, (3) \textit{Fusion} via reciprocal rank fusion
where $k=60$. We sample $K=3$ negatives from top-20 combined ranking (excluding ground truth). This produces challenging multimodal negatives matching keywords and visual appearance without containing correct answers.

\begin{figure}[!h]
  \centering
  \includegraphics[width=\linewidth]{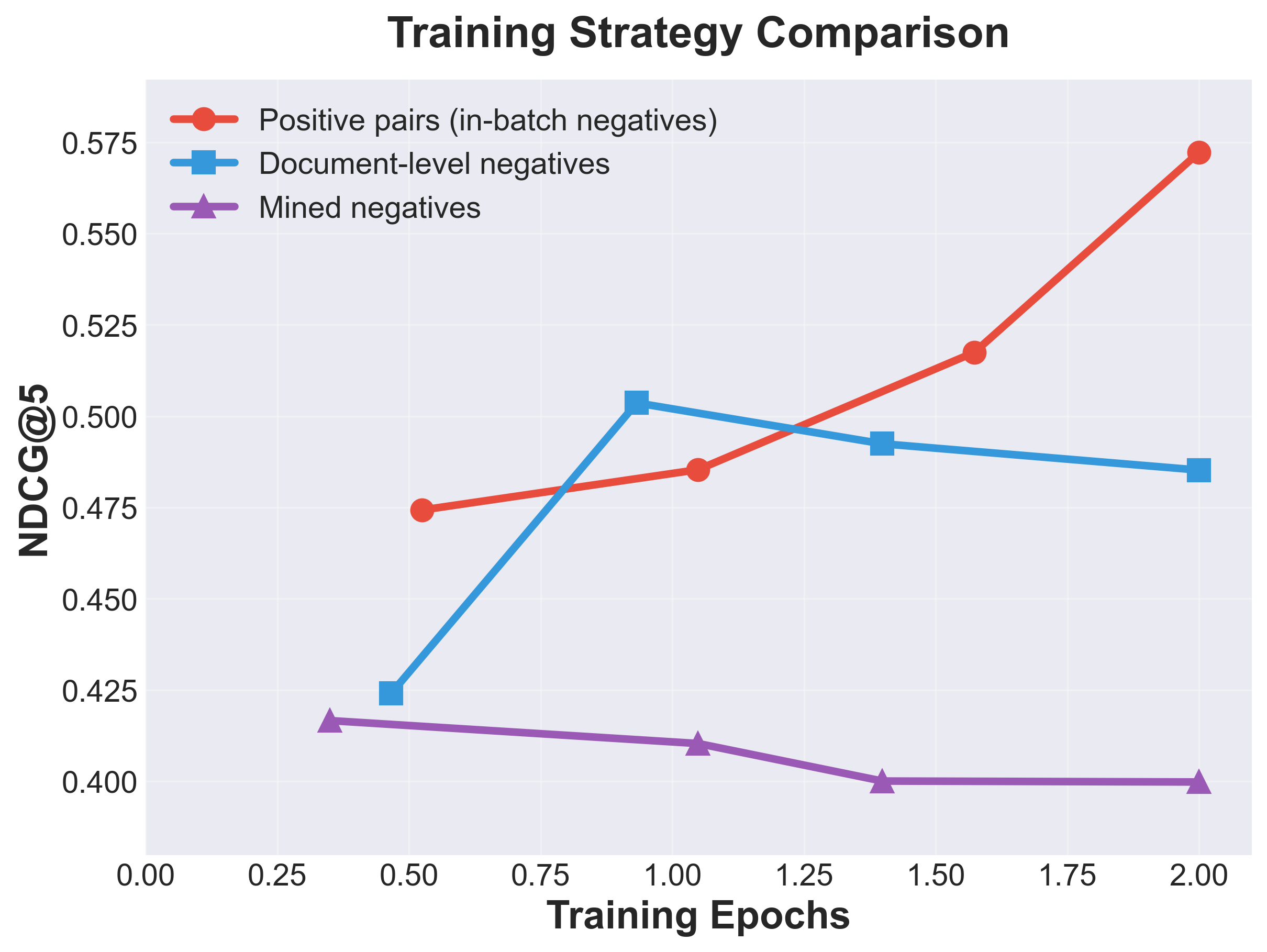}
  \caption{\textbf{Training Strategy Comparison.} Positive-only (in-batch negatives) training strategy substantially outperforms document-level negative and hard negative mining (combined text+visual) strategies, with consistent improvements throughout training.}
  \label{fig:training_comparison}
\end{figure}

\section{Experiments}
\label{sec:experiments}

\subsection{Experimental Setup}
\label{sec:setup}

\noindent\textbf{Baselines.} We compare against state-of-the-art vision-language document retrieval models: ColPali-v1.2~\cite{colpali2024}, ColPali-v1.3, ColQwen2.5-v0.2, ColQwen2-v1.0, ColSmol-500M, GME-Qwen2-VL-2B~\cite{gme2024} jina-embeddings-v4~\cite{jina_v4_2025}, and ColNomic-Embed-Multimodal-3B~\cite{nomicembedmultimodal2025}.

\noindent\textbf{M3DR Models.} We evaluate two 4B-parameter models built on Gemma 3 4B-IT~\cite{gemma3}: \textbf{NetraEmbed} (single dense vector with Matryoshka representation learning) and \textbf{ColNetraEmbed} (ColBERT-style multi-vector following ColPali~\cite{colpali2024}), both trained on 22 languages.

\noindent\textbf{Evaluation Protocol.} We evaluate on two benchmarks: (1) \textit{Nayana-IR} covering 22 languages with both cross-lingual and monolingual retrieval tasks, and (2) \textit{ViDoRe v2}~\cite{vidore2024} for English document retrieval. Primary metric: NDCG@5; secondary metrics: Recall@10, MAP@10, MRR@10.

\subsection{Main Results}
\label{sec:main_results}

Table~\ref{tab:main_results} presents comprehensive evaluation across multilingual and English benchmarks. \textbf{NetraEmbed} reaches 0.716 NDCG@5 on cross-lingual retrieval and 0.738 on monolingual tasks, representing 152\% and 80\% relative improvements over ColPali-v1.3 baseline (0.284 and 0.410 respectively), achieving state-of-the-art multilingual performance. \textbf{ColNetraEmbed} achieves 0.637 NDCG@5 on cross-lingual and 0.670 on monolingual tasks, demonstrating that both paradigms can be effectively trained for multilingual document retrieval, with single dense vectors providing superior efficiency-accuracy trade-offs.

Existing VLM-based retrieval models show substantial performance degradation on multilingual content, with most baselines achieving 0.00-0.32 NDCG@5 on cross-lingual tasks, validating the necessity of explicit multilingual training (\S\ref{appendix:base_model} presents detailed baseline comparisons). On ViDoRe v2 benchmark, NetraEmbed achieves 0.554 NDCG@5, demonstrating competitive performance on English content while prioritizing multilingual capabilities.

\begin{table*}[t]
\centering
\scriptsize
\caption{\textbf{Main Results on Nayana-IR and ViDoRe v2.} M3DR models achieve state-of-the-art multilingual performance while maintaining strong English competitiveness. N@5: NDCG@5, R@10: Recall@10, M@10: MAP@10, MRR: MRR@10.}
\label{tab:main_results}
\setlength{\tabcolsep}{3.5pt}
\begin{tabular*}{\textwidth}{@{\extracolsep{\fill}}l|cccc|cccc|cccc@{}}
\hline
& \multicolumn{4}{c|}{\textbf{Nayana-IR Cross-Lingual}} & \multicolumn{4}{c|}{\textbf{Nayana-IR Monolingual}} & \multicolumn{4}{c}{\textbf{ViDoRe v2}} \\
\textbf{Model} & \textbf{N@5} & \textbf{R@10} & \textbf{M@10} & \textbf{MRR} & \textbf{N@5} & \textbf{R@10} & \textbf{M@10} & \textbf{MRR} & \textbf{N@5} & \textbf{R@10} & \textbf{M@10} & \textbf{MRR} \\
\hline
\multicolumn{13}{l}{\textit{Baselines}} \\
\hline
ColPali-v1.2 & 0.224 & 0.237 & 0.198 & 0.328 & 0.402 & 0.474 & 0.383 & 0.412 & 0.506 & 0.591 & 0.411 & 0.611 \\
ColPali-v1.3 & 0.284 & 0.347 & 0.249 & 0.403 & 0.410 & 0.484 & 0.393 & 0.422 & 0.538 & 0.627 & 0.436 & 0.644 \\
ColQwen2-v1.0 & 0.050 & 0.065 & 0.038 & 0.109 & 0.413 & 0.466 & 0.398 & 0.422 & 0.545 & 0.640 & 0.438 & 0.653 \\
ColQwen2.5-v0.2 & 0.143 & 0.160 & 0.127 & 0.220 & 0.453 & 0.513 & 0.437 & 0.464 & \cellcolor{gray!25}0.592 & 0.664 & \cellcolor{gray!25}0.484 & \cellcolor{gray!25}0.711 \\
ColSmol-500M & 0.000 & 0.000 & 0.000 & 0.000 & 0.224 & 0.263 & 0.214 & 0.229 & 0.435 & 0.544 & 0.347 & 0.535 \\
GME-Qwen2-VL-2B & 0.235 & 0.308 & 0.209 & 0.314 & 0.444 & 0.525 & 0.426 & 0.452 & 0.574 & 0.630 & 0.466 & 0.690 \\
ColNomic-Embed-3B & 0.315 & 0.320 & 0.267 & 0.444 & 0.534 & 0.603 & 0.515 & 0.546 & 0.556 & 0.633 & 0.451 & 0.672 \\
Jina-Embeddings-v4 & 0.435 & 0.435 & 0.390 & 0.548 & X & X & X & X & 0.576 & \cellcolor{gray!25} 0.686 & X & X \\
\hline
\multicolumn{13}{l}{\textit{Our Models}} \\
\hline
\textbf{NetraEmbed} & \cellcolor{gray!25}\textbf{0.716} & \cellcolor{gray!25}\textbf{0.871} & \cellcolor{gray!25}\textbf{0.703} & \cellcolor{gray!25}\textbf{0.775} & \cellcolor{gray!25}\textbf{0.738} & \cellcolor{gray!25}\textbf{0.844} & \cellcolor{gray!25}\textbf{0.709} & \cellcolor{gray!25}\textbf{0.751} & 0.554 & 0.637 & 0.437 & 0.647 \\
\textbf{ColNetraEmbed} & 0.637 & 0.700 & 0.610 & 0.610 & 0.670 & 0.764 & 0.645 & 0.686 & 0.551 & 0.664 & 0.445 & 0.445 \\
\hline
\end{tabular*}
\end{table*}

\subsection{Per-Language Performance Analysis}
\label{sec:language_analysis}

\begin{figure*}[t]
  \centering
  \includegraphics[width=\textwidth]{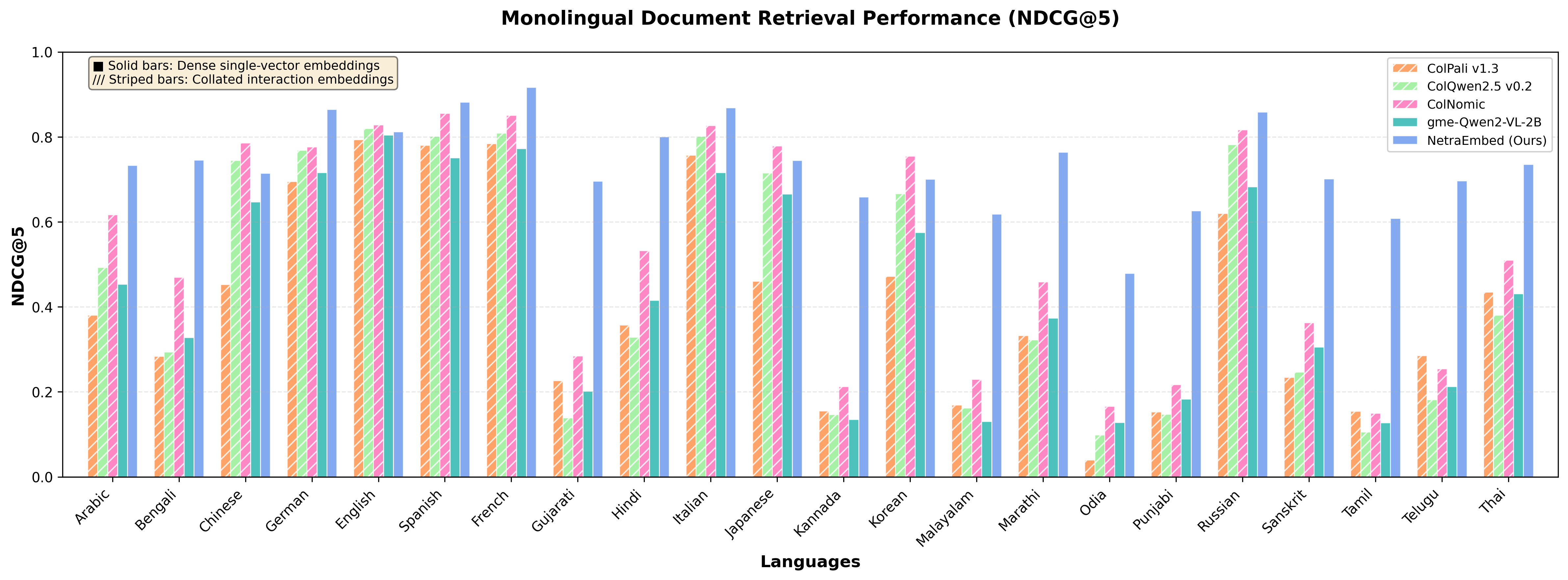}
  \includegraphics[width=\textwidth]{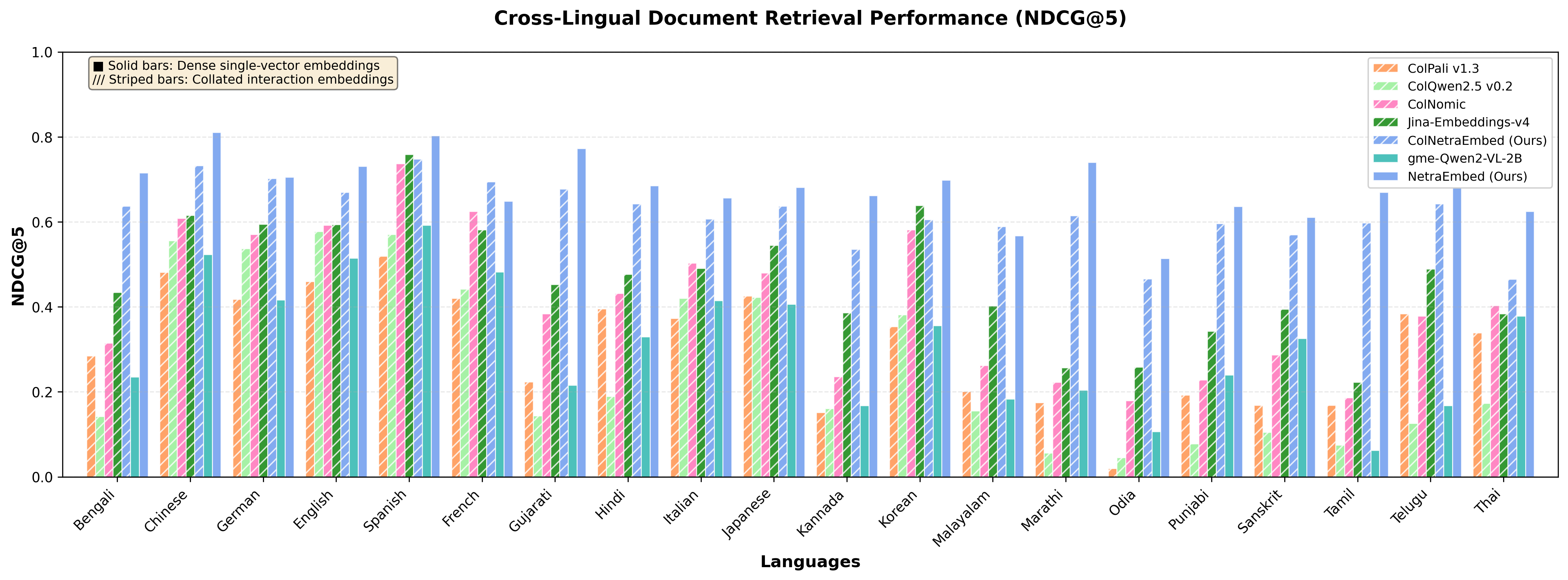}
  \caption{\textbf{Per Language Performance Across 22 Languages.} NetraEmbed achieves consistent high performance across all languages and script families such as Latin, Devanagari, CJK, Arabic, and others, while English centric baselines show significant drops on non English content.}
  \label{fig:language_performance}
\end{figure*}

Figure~\ref{fig:language_performance} illustrates per-language performance across all 22 languages in our benchmark. NetraEmbed achieves consistent high performance across diverse script families, validating robust cross-lingual transfer. Scaling analysis from 6 to 22 languages (\S\ref{appendix:scaling}) demonstrates progressive improvements as linguistic diversity increases.

\subsection{Matryoshka Embeddings: Efficiency-Accuracy Trade-offs}
\label{sec:matryoshka_analysis}

\begin{table}[H]
\centering
\small
\caption{\textbf{Matryoshka Embedding Dimension Trade offs.} Storage and performance for different truncation levels of NetraEmbed on Nayana IR cross lingual benchmark.}
\label{tab:matryoshka}
\resizebox{0.45\textwidth}{!}{%
\setlength{\tabcolsep}{10pt}
\begin{tabular}{lccc}
\hline
\textbf{Dimensions} & \textbf{Storage/Doc} & \textbf{NDCG@5} & \textbf{Rel. Perf.} \\
\hline
768 & $\sim$3 KB & 0.680 & 95.0\% \\
1536 & $\sim$6 KB & 0.706 & 98.6\% \\
2560 (full) & $\sim$10 KB & 0.716 & 100.0\% \\
\hline
\end{tabular}%
}
\end{table}

Matryoshka representation learning enables flexible post-deployment dimension selection without retraining, critical for adapting to diverse computational budgets. Table~\ref{tab:matryoshka} quantifies storage-accuracy trade-offs for NetraEmbed. Comprehensive ablations across training configurations and language scales are presented in \S\ref{appendix:matryoshka}.

The 768-dimensional truncation achieves 95.0\% of full model performance (0.680 vs 0.716 NDCG@5) while reducing storage by 70\% ($\sim$3 KB vs $\sim$10 KB per document), making it ideal for billion-scale deployments or edge devices with strict memory constraints. The 1536-dimensional representation offers a balanced middle ground at 98.6\% performance with 40\% storage reduction, suitable for production systems balancing accuracy and cost. The full 2560-dimensional representation maximizes retrieval accuracy while maintaining efficient storage requirements compared to token-level approaches, making NetraEmbed suitable for large-scale multilingual document retrieval deployments.




\subsection{Single Dense Vector vs. ColBERT-Style Multi-Vector Retrieval}
\label{sec:dense_vs_late}

M3DR successfully generalizes across both retrieval paradigms. We compare the two approaches to understand architectural trade-offs for multilingual document retrieval.

\medskip

\noindent\textbf{Performance.} NetraEmbed consistently outperforms ColNetraEmbed across multilingual tasks: +12.4\% on cross-lingual NDCG@5 (0.716 vs 0.637) and +10.1\% on monolingual tasks (0.738 vs 0.670). Both models achieve similar English performance on ViDoRe v2 (0.554 vs 0.551), suggesting that the single dense vector approach particularly benefits cross-lingual transfer learning.

\medskip

\noindent\textbf{Efficiency.} Single dense vectors offer substantial computational advantages: (1) \textit{Storage}: $\sim$10 KB per document vs. $\sim$2.5 MB for ColBERT-style multi-vector with 256 image tokens (250$\times$ reduction), (2) \textit{Retrieval Speed}: Single-vector cosine similarity via HNSW vs. expensive MaxSim computation over token matrices, (3) \textit{Scalability}: Single dense vectors scale efficiently to billion-document corpora with standard vector databases. \\


Our results demonstrate that M3DR successfully generalizes across both paradigms, with single dense vectors providing the optimal balance of accuracy, efficiency, and multilingual generalization for large-scale document retrieval deployments. Detailed architectural comparisons and deployment considerations are provided in \S\ref{appendix:dense_vs_col}.

\section{Conclusion}
\label{sec:conclusion}

This work addresses a fundamental gap in vision-based document retrieval: the inability of existing systems to handle multilingual content effectively. We presented M3DR, a comprehensive framework that achieves robust multilingual multimodal document retrieval across 22 typologically diverse languages spanning Latin, Devanagari, Dravidian, CJK, Arabic, and other script families.

\medskip

\noindent\textbf{Key Results.} NetraEmbed achieves 0.716 NDCG@5 on cross-lingual retrieval and 0.738 on monolingual tasks, representing 152\% and 80\% relative improvements over the strongest baseline (ColPali-v1.3: 0.284 and 0.410 respectively). These gains demonstrate that explicit multilingual training is essential existing English-centric VLMs catastrophically fail on non-English content despite strong English performance. NetraEmbed maintains competitive English performance (0.554 NDCG@5 on ViDoRe v2) while dramatically improving multilingual capabilities, validating that cross-lingual optimization does not require sacrificing monolingual competitiveness.

\medskip

\noindent\textbf{Critical Design Insights.} Through extensive ablations (\S\ref{appendix:overview}--\ref{appendix:training_config}), we identified several key principles: (1) \textit{Base model selection is decisive} Gemma 3 4B-IT~\cite{gemma3}'s multilingual pretraining enabled significant point gains over English centric models (ColQwen2, ColPali) on cross lingual tasks. (2) \textit{Training simplicity wins} BiEncoderLoss with in batch negatives outperformed complex hard negative mining strategies, as multilingual batch diversity provides sufficient contrastive signal. (3) \textit{Last token pooling dominates} outperforming mean pooling by 13+ NDCG@5 points for decoder based VLMs. (4) \textit{Matryoshka enables flexibility} 768 dimensional truncation retains 95\% performance with 70\% storage reduction.

\medskip

\noindent\textbf{Resources and Reproducibility.} We release the Nayana-IR benchmark (23 datasets, $\sim$28K images, $\sim$5.4K queries in BEIR format) covering cross-lingual and monolingual retrieval across 22 languages. Our models NetraEmbed and ColNetraEmbed are publicly available, alongside training code and evaluation scripts. With LoRA fine-tuning on 4$\times$A100 GPUs requiring only $\sim$12 hours for SOTA results, we aim to democratize multilingual document retrieval research for groups with modest computational resources.

\medskip

\noindent\textbf{Broader Impact.} M3DR democratizes document intelligence across linguistic communities, enabling equitable access to information retrieval technologies regardless of language. Applications span multilingual enterprise knowledge management, cross-border research collaboration, educational resource discovery for underserved languages, and cultural heritage preservation for digitized manuscripts. By achieving state-of-the-art multilingual performance without sacrificing English competitiveness, M3DR demonstrates that inclusive AI systems need not compromise on quality. However, practitioners must monitor performance disparities across languages in deployment and implement continuous evaluation and mitigation strategies to ensure fairness and prevent marginalization of lower-resource languages.

\medskip

\noindent\textbf{Limitations and Future Directions.} While our results demonstrate strong generalization across 22 languages, several limitations warrant further investigation: (1) Performance on rare language pairs (e.g., Tamil$\rightarrow$Russian) shows 10-12\% degradation compared to high-resource pairs, suggesting opportunities for improved cross-lingual alignment techniques. (2) Complex tabular content with language-specific number formatting (Hindi numerals, Arabic-Indic digits) remains challenging, indicating potential benefits from table-aware training objectives. (3) Our evaluation focuses on document-level retrieval; extending to passage-level or region-level retrieval within documents presents interesting future work. (4) Scaling beyond 22 languages to truly low-resource languages and investigating zero-shot transfer to unseen scripts would further validate the framework's generalization capacity.

\section*{Acknowledgments}
This work benefited from compute credits for training, inference, and evaluation provided by \href{https://modal.com}{Modal}, acknowledged as a compute sponsor. Dataset curation and synthesis were supported by the \href{https://about.fb.com/news/2025/04/llama-impact-grant-recipients/?utm_source=AIatMeta&utm_medium=organic_social&utm_content=image&utm_campaign=llamacon}{Meta LLaMA Impact Grant} through our \href{https://www.cognitivelab.in/nayana}{Nayana initiative}. We appreciate Meta for continued support of our research efforts at \href{https://www.cognitivelab.in}{CognitiveLab}.

\clearpage

{
    \small
    \bibliographystyle{ieeenat_fullname}
    \bibliography{main}
}

\onecolumn
\appendix

\section*{Appendix}
\addcontentsline{toc}{section}{Appendix}

\section{Overview}
\label{appendix:overview}

This appendix presents a comprehensive analysis of the ablation studies conducted to develop the M3DR (Multilingual Multimodal Dense Retrieval) framework. We initially ran small ablations for each method with a dataset of 45k images across 6 languages to determine if it made sense to run larger scale experiments. Based on these preliminary results, we progressively scaled up to 22 languages with approximately 250k image text pairs.

\medskip

The ablations cover critical design decisions including:
\begin{itemize}
    \item Base model selection
    \item Loss function comparisons
    \item Pooling strategies
    \item Matryoshka embedding dimensions
    \item Model merging techniques
    \item Dense vs. late interaction architectures
    \item Cross-lingual scaling behavior
\end{itemize}
\medskip
All experiments were evaluated on three benchmarks:
\begin{itemize}
    \item \textbf{ViDoRe v2}: Document retrieval across 4 languages (English, French, German, Spanish)
    \item \textbf{NayanaIR-CrossBench}: Cross-lingual retrieval across 20 languages
    \item \textbf{NayanaIR-Bench-Monolingual}: Monolingual retrieval evaluation (22 datasets, 1 for each language)
\end{itemize}

\noindent \textbf{Metrics Reported}: NDCG@5 (N@5), Recall@10 (R@10), MAP@10 (M@10), MRR@10 (MRR) -- consistent with main paper results.

\begin{tcolorbox}[colback=gray!10, colframe=black!50, title=Model Naming Conventions]
\textbf{Architecture Types:}
\begin{itemize}
    \item \textbf{SV (Single-Vector)}: Models producing one dense embedding per image/query for efficient similarity search
    \item \textbf{MV (Multi-Vector)}: ColBERT-style late interaction models producing multiple embeddings per image/query with MaxSim matching
\end{itemize}
\textbf{Base Models:}
\begin{itemize}
    \item \textbf{Gemma3}: Our models built on Gemma 3 4B~\cite{gemma3} backbone (finetuned versions)
    \item \textbf{colpali/colqwen/etc.}: External baseline models (original names preserved)
\end{itemize}
\end{tcolorbox}

\FloatBarrier
\section{Base Model Selection}
\label{appendix:base_model}

This ablation study helped us decide which base model to use for our framework. We evaluated existing pre-trained models using two different retrieval architectures:

\begin{itemize}
    \item \textbf{Col (Late Interaction)}: Multi-vector ColBERT-style models that produce multiple embeddings per image/query and use late interaction (MaxSim) for retrieval (e.g., ColPali, ColQwen, ColSmol)
    \item \textbf{Dense}: Single-vector embedding models that produce one dense embedding per image/query for efficient vector search (e.g., Gemma3, GME-Qwen, Colnomic)
\end{itemize}
\medskip
While existing Col models performed well on ViDoRe v2, they were significantly behind on the NayanaIR-CrossBench, indicating poor cross-lingual generalization. We also tested an initial dense embedding model based on Gemma3 4B (Gemma3-InBatch) trained on the ColPali training set with in-batch negative loss.
\medskip
\begin{table}[!htbp]
\centering
\caption{\textbf{Base Model Selection: Performance Across Benchmarks.} Comparison reveals catastrophic cross-lingual failure of models pretrained primarily on English data. SV = Single-Vector; MV = Multi-Vector.}
\label{tab:base_model_comparison_full}
\resizebox{\textwidth}{!}{%
\begin{tabular}{p{5cm}ccccccc}
\hline
\textbf{Model} & \textbf{Type} & \textbf{ViDoRe N@5} & \textbf{ViDoRe R@10} & \textbf{Cross N@5} & \textbf{Cross R@10} & \textbf{Mono N@5} & \textbf{Mono R@10} \\
\hline
colqwen2.5-v0.2 & MV & 59.20 & 66.44 & \textcolor{red}{14.26} & 15.98 & 45.27 & 51.30 \\
ColQwen2-6langs & MV & 58.63 & 67.62 & 37.15 & 39.85 & N/A & N/A \\
gme-Qwen2-VL-2B & SV & 57.39 & 63.01 & 23.53 & 30.83 & 44.36 & 52.51 \\
ColPali-6langs & MV & 56.41 & 64.94 & 42.22 & 50.05 & 53.38 & 62.34 \\
colnomic-embed-3b & SV & 55.55 & 63.26 & 31.47 & 31.96 & 53.37 & 60.28 \\
colqwen2-v1.0 & MV & 54.54 & 64.04 & \textcolor{red}{4.97} & 6.47 & 41.27 & 46.58 \\
colpali-v1.3 & MV & 53.85 & 62.69 & 28.45 & 34.71 & 41.03 & 48.41 \\
Gemma3-InBatch & SV & 52.44 & 60.34 & 20.65 & 23.14 & 37.95 & 46.91 \\
colpali-v1.2 & MV & 50.55 & 59.07 & 22.41 & 23.68 & 40.21 & 47.45 \\
colSmol-500M & MV & 43.49 & 54.41 & \textcolor{red}{0.00} & 0.00 & 22.41 & 26.26 \\
\hline
\end{tabular}%
}
\end{table}

\begin{figure}[!htbp]
\centering
\includegraphics[width=0.9\textwidth]{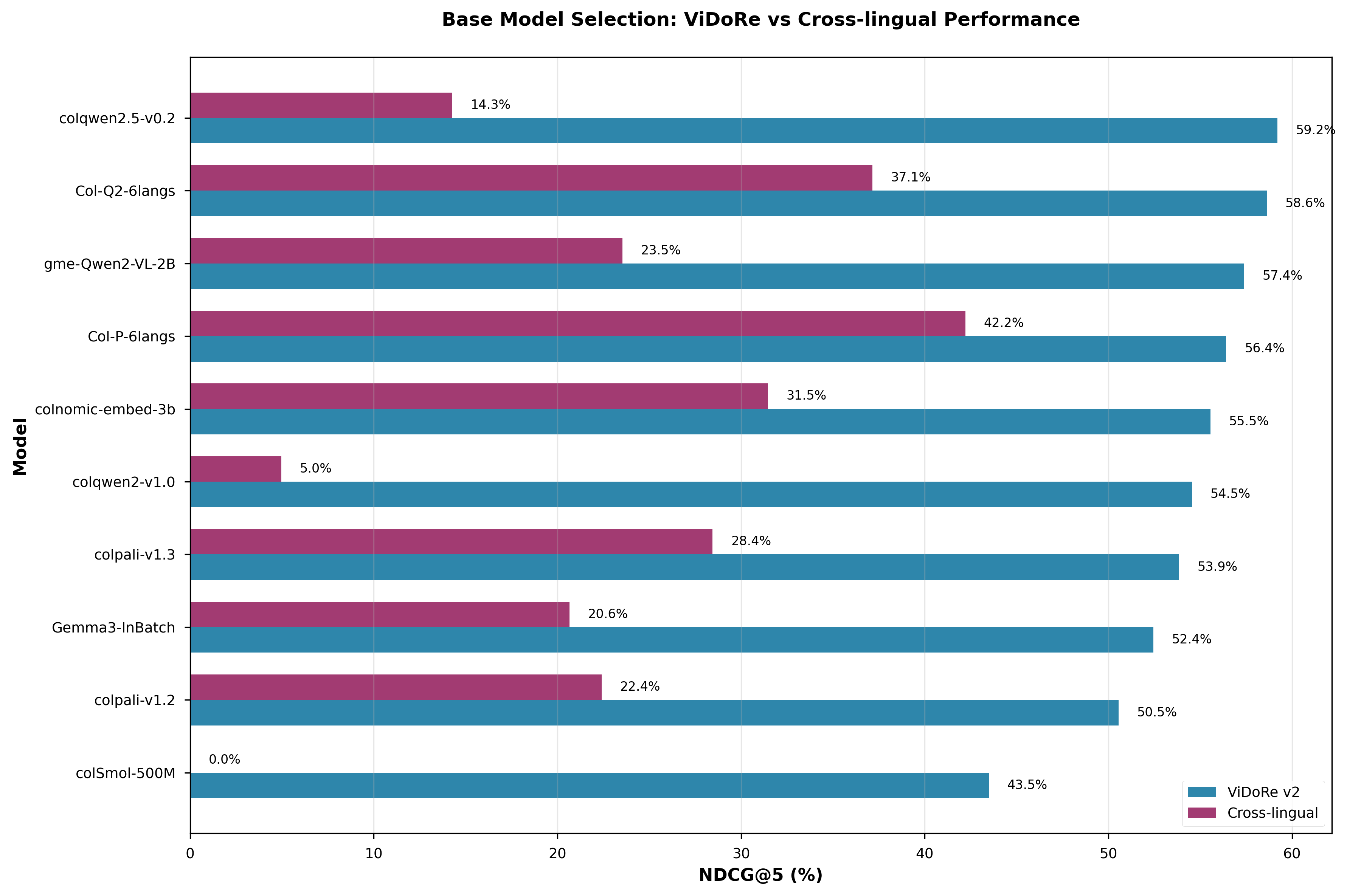}
\caption{\textbf{Base Model Comparison: ViDoRe vs Cross-lingual NDCG@5 for all baseline models.} Models achieving high ViDoRe performance (English-dominated) often fail catastrophically on cross-lingual tasks.}
\label{fig:base_model_comparison}
\end{figure}

Figure~\ref{fig:base_model_comparison} visualizes the stark trade-off between ViDoRe and cross-lingual performance across baseline models. The scatter plot reveals a concerning pattern: models that achieve top performance on English-centric ViDoRe v2 (such as colqwen2.5-v0.2 at 59.20\% NDCG@5) experience catastrophic failure on cross-lingual tasks, plummeting to just 14.26\% NDCG@5, a 45-\%-point drop. This visualization demonstrates that despite using single-vector dense embeddings, our initial Gemma3-InBatch model shows promise due to Gemma3 4B's robust multilingual vocabulary and pretraining. The figure clearly illustrates why we selected Gemma3 4B as our foundation: its multilingual capacity enables cross-lingual transfer that English-centric ColPali and ColQwen2 architectures cannot match, even when these models dominate on document retrieval benchmarks.

\FloatBarrier
\section{Preliminary Ablations on 6 Languages}
\label{appendix:6lang_ablations}

Before committing substantial compute to full-scale experiments, we conducted targeted ablations on a smaller 6-language subset (Hindi, Kannada, Tamil, Telugu, Chinese, Japanese) comprising $\sim$45k image-text pairs.

\FloatBarrier
\subsection{Loss Function Ablations}

We compared different loss functions for both dense and Col models to determine the optimal training objective. For dense models, we tested BiEncoderLoss, BiNegativeCELoss, BiPairwiseCELoss, and BiPairwiseNegativeCELoss. For Col models, we compared ColBERT loss against ColBERT with pairwise loss.

\subsubsection{Dense Model Loss Functions}

We evaluated BiEncoderLoss (InBatch) as the baseline and compared it against hard negative mining at different training checkpoints to understand how hard negative mining affects model convergence over time.

\begin{table}[!htbp]
\centering
\caption{\textbf{Dense Model Loss Function Ablation on ViDoRe v2.}}
\label{tab:dense_loss_ablation_detailed}
\begin{tabular}{lccccc}
\hline
\textbf{Model} & \textbf{Loss Type} & \textbf{NDCG@5} & \textbf{Recall@10} & \textbf{MAP@10} & \textbf{MRR@10} \\
\hline
Gemma3-InBatch (ckpt-2000) & BiEncoderLoss & 49.31 & 58.14 & 38.78 & 59.03 \\
Gemma3-HardNeg (ckpt-750) & Hard Negative & 50.48 & 60.22 & 39.72 & 61.24 \\
Gemma3-HardNeg (ckpt-1500) & Hard Negative & 49.23 & 55.72 & 38.26 & 60.54 \\
Gemma3-HardNeg (ckpt-1694) & Hard Negative & 50.12 & 62.03 & 39.26 & 59.86 \\
Gemma3-HardNeg (ckpt-1950) & Hard Negative & 49.31 & 58.80 & 38.49 & 59.91 \\
Gemma3-HardNeg (ckpt-2145) & Hard Negative & 49.27 & 56.84 & 39.11 & 60.49 \\
Gemma3-HardNeg (ckpt-2300) & Hard Negative & 46.73 & 59.22 & 37.68 & 56.25 \\
\hline
\end{tabular}%
\end{table}

\begin{figure}[!htbp]
\centering
\includegraphics[width=0.8\textwidth]{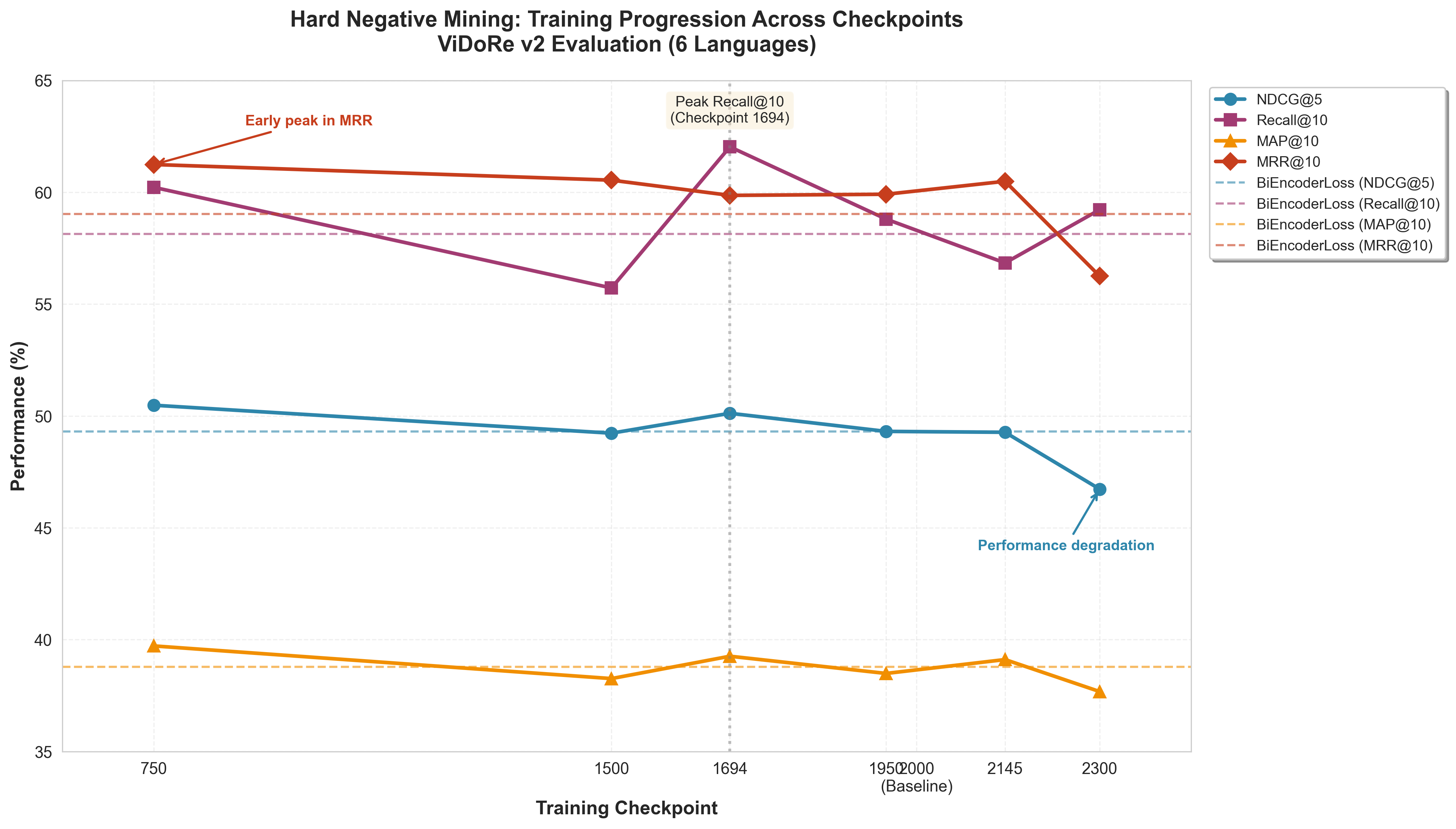}
\caption{\textbf{Hard Negative Mining Training Progression.} Line chart showing metric evolution across training steps (750-2300).}
\label{fig:hardneg_training_progression}
\end{figure}

Figure~\ref{fig:hardneg_training_progression} traces the evolution of four key metrics (NDCG@5, Recall@10, MAP@10, MRR@10) across training checkpoints when using hard negative mining. The line chart reveals erratic behavior: NDCG@5 fluctuates between 46.73\% and 50.48\% without clear convergence, while Recall@10 peaks at checkpoint 1694 (62.03\%) before declining. This volatility contrasts sharply with the stable convergence of standard BiEncoderLoss, suggesting that hard negative mining introduces training instability in our multilingual multimodal setting. The visualization demonstrates that while hard negative mining provided marginal peak performance gains (1-3 points), the instability and careful checkpoint selection required make it impractical compared to the simpler in-batch negative approach.

\subsubsection{Col Model Loss Functions}

\begin{table}[!htbp]
\centering
\caption{\textbf{Col Model Loss Function Ablation.}}
\label{tab:col_loss_ablation}
\begin{tabular}{lcccc}
\hline
\textbf{Model} & \textbf{Loss} & \textbf{ViDoRe NDCG@5} & \textbf{ViDoRe Recall@10} & \textbf{Cross NDCG@5} \\
\hline
ColPali-6langs-ColBERT & ColBERT & 56.41 & 64.94 & 42.22 \\
ColPali-6langs-Pairwise & ColBERT+Pairwise & 39.53 & 51.98 & 33.94 \\
\hline
\end{tabular}%
\end{table}

\begin{figure}[!htbp]
\centering
\includegraphics[width=0.7\textwidth]{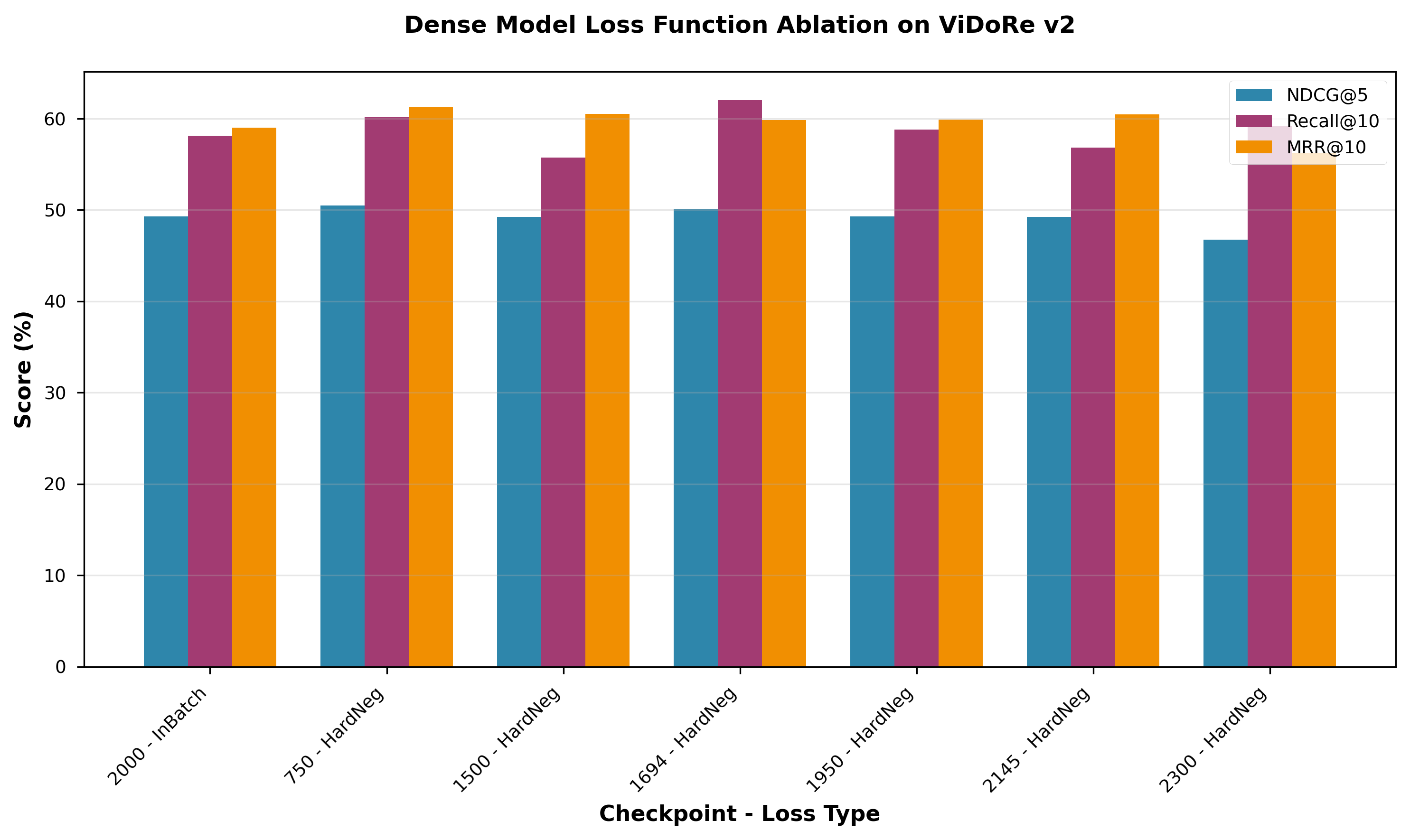}
\caption{\textbf{Dense Model Loss Function Ablation.} Comparing BiEncoderLoss and Hard Negative mining variants.}
\label{fig:dense_loss_ablation}
\end{figure}

The comparison reveals that for single-vector dense models, BiEncoderLoss emerged as the optimal choice, while for multi-vector Col models, standard ColBERT loss significantly outperformed ColBERT with pairwise loss (56.41\% vs 39.53\% NDCG@5 on ViDoRe), a 17 percentage point degradation. This is primarily because both the ColPali train set and Nayana IR dataset contained positive pairs without explicitly annotated hard negatives, making the diversity provided by in-batch negatives across languages and visual content sufficient for learning discriminative representations. Figure~\ref{fig:dense_loss_ablation} summarizes these findings, illustrating that simpler objectives outperform complex negative sampling in multilingual settings.

\FloatBarrier
\subsection{Pooling Strategy Ablations: Last Token vs Mean Pooling}

For single-vector dense embedding models, we need to convert the final hidden states into a single embedding vector. Note: This ablation is specific to dense models only, as Col models inherently use all patch embeddings without pooling.

\begin{table}[!htbp]
\centering
\caption{\textbf{Pooling Strategy Ablation on ViDoRe v2.}}
\label{tab:pooling_ablation_detailed}
\begin{tabular}{lcccc}
\hline
\textbf{Model} & \textbf{Pooling} & \textbf{NDCG@5} & \textbf{Recall@10} & \textbf{MRR@10} \\
\hline
Gemma3-LastToken & Last Token & \textbf{49.31} & \textbf{58.14} & \textbf{59.03} \\
Gemma3-MeanPool & Mean Pooling & 35.85 & 45.04 & 45.07 \\
\hline
\end{tabular}%
\end{table}

\begin{figure}[!htbp]
\centering
\includegraphics[width=0.6\textwidth]{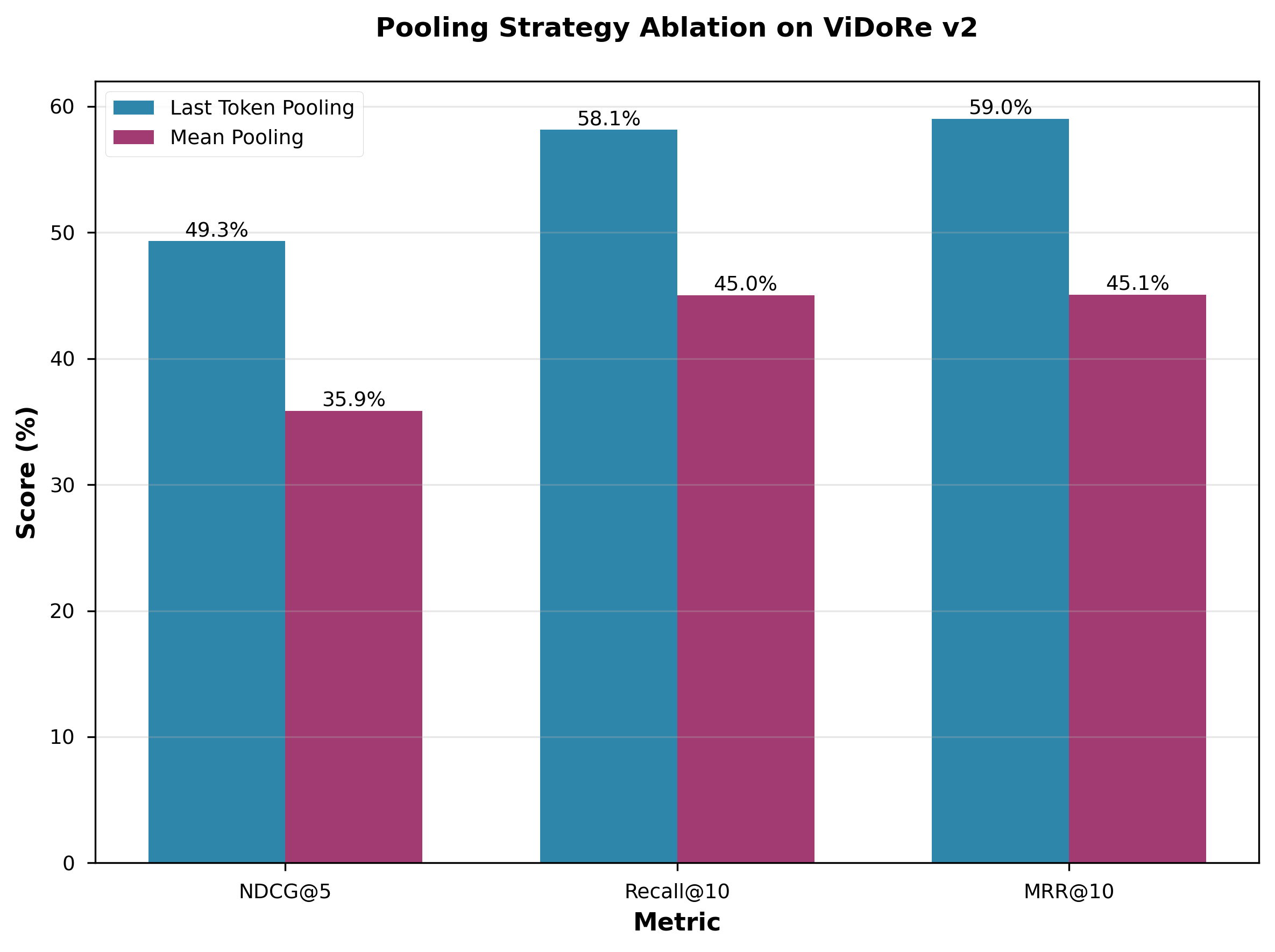}
\caption{\textbf{Pooling Strategy Ablation.} Last Token vs Mean Pooling performance comparison.}
\label{fig:pooling_ablation}
\end{figure}

Figure~\ref{fig:pooling_ablation} demonstrates the dramatic superiority of last token pooling over mean pooling, with performance differences of 13.5 percentage points in NDCG@5 (49.31\% vs 35.85\%). This suggests that decoder-only VLMs like Gemma encode critical summary information in the final token representation, similar to how GPT-style models aggregate sequence information. Additionally, last token pooling proves computationally cheaper during inference as it requires processing only a single token representation rather than averaging across all tokens, avoiding the dilution of salient information through averaging. This became our default pooling strategy for all dense embedding models.

\FloatBarrier
\subsection{OCR Model Ablations}

We hypothesized that existing VLMs pre-trained for OCR tasks (such as Docling and Granite OCR) might generalize well to text-heavy document retrieval. We finetuned these models for both dense and Col-style retrieval to test this hypothesis.

\begin{table}[!htbp]
\centering
\caption{\textbf{OCR Model Transfer Ablation on ViDoRe v2.} SV = Single-Vector; MV = Multi-Vector.}
\label{tab:ocr_ablation_detailed}
\begin{tabular}{lcccc}
\hline
\textbf{Model} & \textbf{Type} & \textbf{NDCG@5} & \textbf{Recall@10} & \textbf{MRR@10} \\
\hline
Granite-Docling-258M & OCR baseline & 0.97 & 1.16 & 2.40 \\
Docling-Finetuned & SV & 4.17 & 6.92 & 5.72 \\
Gemma3-OCRInit & MV & 1.37 & 1.05 & 1.94 \\
Gemma3-Pretrained & VLM (no FT) & 4.48 & 3.43 & 8.52 \\
\hline
\end{tabular}%
\end{table}

OCR-pretrained models failed to generalize to retrieval tasks even after finetuning. The pretrained Granite-Docling achieved near-zero retrieval performance (0.97\% NDCG@5), and finetuning only recovered to 4.17\%, still catastrophically poor. Remarkably, the base Gemma 3 4B~\cite{gemma3} model without any finetuning matched this performance. This negative result demonstrates that OCR pretraining optimizes for character-level text recognition and spatial layout understanding, creating representations specialized for transcription rather than semantic retrieval. These representations are orthogonal to the semantic similarity space required for retrieval, validating our choice of general-purpose VLMs over task-specific OCR models.

\FloatBarrier
\subsection{Training on 6 Languages: Initial Results}

We finetuned existing multi-vector Col models (ColQwen2 and ColPali) and trained single-vector dense Gemma3~\cite{gemma3} models on our 6-language dataset to evaluate cross-lingual generalization capabilities.

\begin{table}[!htbp]
\centering
\caption{\textbf{6-Language Training Results.} SV = Single-Vector; MV = Multi-Vector.}
\label{tab:6lang_results_detailed}
\resizebox{\textwidth}{!}{%
\begin{tabular}{p{4.5cm}ccccccc}
\hline
\textbf{Model} & \textbf{Type} & \textbf{ViDoRe N@5} & \textbf{ViDoRe R@10} & \textbf{Cross N@5} & \textbf{Cross R@10} & \textbf{Mono N@5} & \textbf{Mono R@10} \\
\hline
ColQwen2-6langs & MV & 58.63 & 67.62 & 37.15 & 39.85 & N/A & N/A \\
ColPali-6langs & MV & 56.41 & 64.94 & 42.22 & 50.05 & 53.38 & 62.34 \\
Gemma3-6langs & SV & 49.08 & 59.76 & 60.39 & 72.94 & 62.48 & 74.70 \\
Gemma3-6langs-v2 & SV & 49.06 & 60.44 & 59.62 & 76.47 & 62.42 & 73.88 \\
\hline
\end{tabular}%
}
\end{table}

\begin{figure}[!htbp]
\centering
\includegraphics[width=0.8\textwidth]{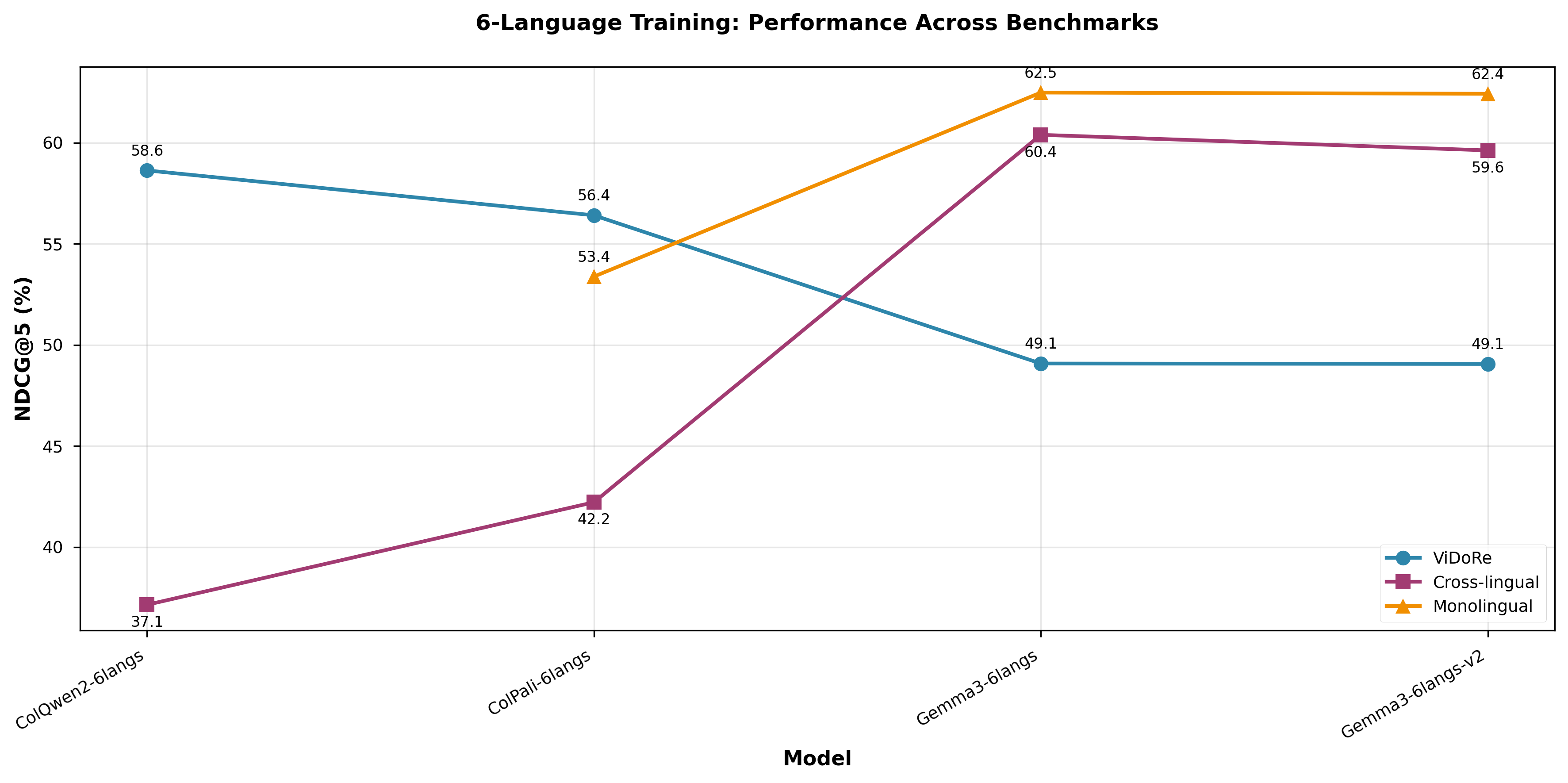}
\caption{\textbf{6-Language Training: Performance across benchmarks.} Dense Gemma3 models excel at cross-lingual retrieval.}
\label{fig:6lang_comparison}
\end{figure}

Figure~\ref{fig:6lang_comparison} presents a pivotal finding in our research journey. While multi-vector ColQwen2 and ColPali variants maintained their ViDoRe advantage (56-59\% vs 49\% NDCG@5), the visualization reveals that dense Gemma3 models achieved dramatically superior cross-lingual performance: 60.39\% vs 42.22\% for the best late interaction model, an 18 \% point gap. This massive difference validates our base model hypothesis: Gemma's multilingual pretraining enabled cross-lingual transfer that English-centric ColPali and ColQwen2 foundations could not match, even after multilingual finetuning. The figure demonstrates the clear superiority of Gemma3-based models across cross-lingual and monolingual benchmarks, convincing us to pursue both single-vector Gemma3 as the primary model and multi-vector ColGemma3 for a comprehensive architectural comparison.

\FloatBarrier
\section{Matryoshka Embedding Ablations}
\label{appendix:matryoshka}

Matryoshka Representation Learning allows single-vector dense models to produce embeddings that maintain performance across different dimensionalities. This technique is specific to dense embedding models and does not apply to multi-vector Col models.

\FloatBarrier
\subsection{Matryoshka on 6 Languages}

\begin{table}[!htbp]
\centering
\caption{\textbf{Matryoshka Ablation on 6 Languages.}}
\label{tab:matryoshka_6lang_detailed}
\begin{tabular}{lccccc}
\hline
\textbf{Model} & \textbf{Dimensions} & \textbf{ViDoRe N@5} & \textbf{ViDoRe R@10} & \textbf{Cross N@5} & \textbf{Cross R@10} \\
\hline
Gemma3-Matryoshka-6langs & 2056 & 45.81 & 60.32 & 57.22 & 71.32 \\
Gemma3-6langs (baseline) & 2056 & 49.08 & 59.76 & 60.39 & 72.94 \\
\hline
\end{tabular}%
\end{table}

The 6-language Matryoshka model showed 3-point degradation on ViDoRe and Cross benchmarks, with slight improvement on monolingual tasks. While concerning, we hypothesized that scaling to more languages might provide sufficient training signal to overcome this gap.

\FloatBarrier
\subsection{Matryoshka on 22 Languages (Final Scale)}

These models were trained on the ColPali train set and then finetuned with Matryoshka loss [768, 1536, 2056] and BiNegative CE loss on the Nayana IR dataset.

\begin{table}[!htbp]
\centering
\caption{\textbf{Primary Matryoshka Models on 22 Languages.}}
\label{tab:matryoshka_22lang_primary}
\resizebox{\textwidth}{!}{%
\begin{tabular}{lcccccccc}
\hline
\textbf{Model} & \textbf{Dimensions} & \textbf{ViDoRe N@5} & \textbf{ViDoRe R@10} & \textbf{Cross N@5} & \textbf{Cross R@10} & \textbf{Mono N@5} & \textbf{Mono R@10} \\
\hline
Gemma3-Matryoshka (2056) & 2056 & 49.88 & 62.00 & 77.31 & 88.38 & 74.10 & 85.03 \\
Gemma3-Matryoshka (1536) & 1536 & 48.21 & 60.47 & 76.87 & 86.91 & 73.25 & 84.07 \\
Gemma3-Matryoshka (768) & 768 & 45.10 & 56.52 & 73.15 & 79.12 & 70.77 & 82.42 \\
\hline
Gemma3-Matryoshka-v2 (2056) & 2056 & 45.63 & 57.33 & 72.60 & 83.09 & 73.77 & 84.29 \\
Gemma3-Matryoshka-v2 (1536) & 1536 & 45.00 & 56.42 & 72.16 & 82.50 & 73.14 & 84.02 \\
Gemma3-Matryoshka-v2 (768) & 768 & 42.86 & 53.91 & 70.45 & 84.12 & 72.05 & 82.68 \\
\hline
\end{tabular}%
}
\end{table}

\begin{figure}[!htbp]
\centering
\includegraphics[width=0.8\textwidth]{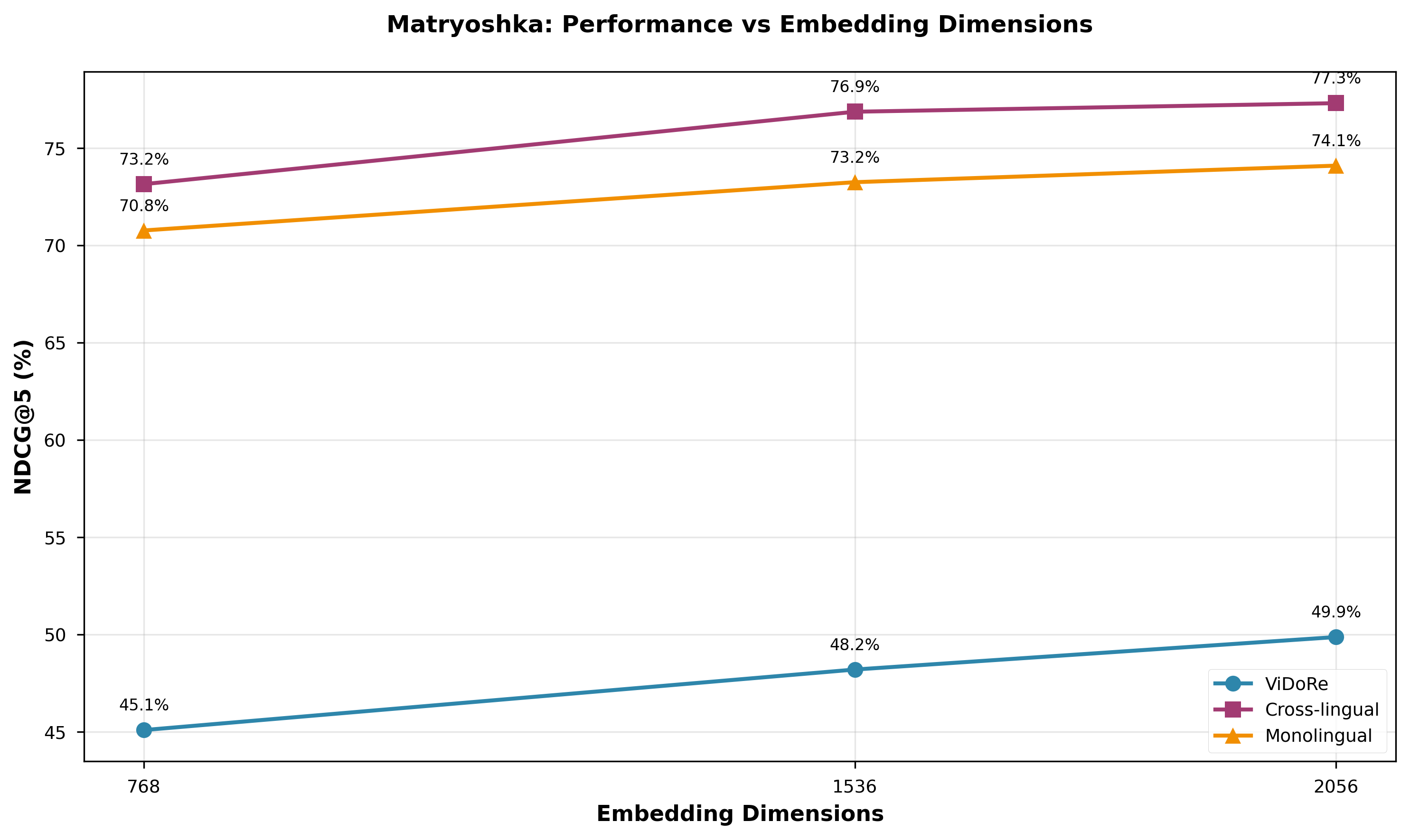}
\caption{\textbf{Matryoshka Embedding Dimensions.} NDCG@5 performance across 768, 1536, and 2056 dimensions showing graceful degradation.}
\label{fig:matryoshka_dimensions}
\end{figure}

Figure~\ref{fig:matryoshka_dimensions} illustrates the graceful degradation of performance as embedding dimensionality decreases. The visualization shows that 768-dimensional embeddings retained 94.6\% of full performance on cross-lingual tasks (73.15\% vs 77.31\% NDCG@5), a mere 4.2-point drop, while providing 70\% storage reduction. The 1536-dimensional embeddings achieved 99.4\% of full performance (76.87\% vs 77.31\%), effectively trading 1 NDCG@5 point for 40\% storage savings. This flexibility is critical for production systems, where index size directly impacts cost and latency. Both the primary and v2 training runs showed consistent behavior across dimensions, confirming the robustness of Matryoshka training and enabling deployment teams to choose the optimal performance-storage trade-off for their specific use case.

\FloatBarrier
\section{Model Merging Strategies}
\label{appendix:merging}

To create a more balanced model that performs well across both ViDoRe and Nayana cross-lingual benchmarks, we explored model merging techniques. We merged two complementary checkpoints: Parent Model A (Gemma3-CrossLingual) with excellent cross-lingual performance (77.31\% NDCG@5) but moderate ViDoRe performance (49.88\%), and Parent Model B (Gemma3-ViDoRe) with top ViDoRe performance (55.40\%) but decent cross-lingual performance (71.57\%).

We explored two interpolation methods: Linear merging (weighted average $\alpha \cdot \theta_1 + (1-\alpha) \cdot \theta_2$) and SLERP (Spherical Linear Interpolation, interpolating along the geodesic on the hypersphere).
\medskip
\begin{table}[!htbp]
\centering
\caption{\textbf{Model Merging Results.}}
\label{tab:model_merging_full}
\resizebox{\textwidth}{!}{%
\begin{tabular}{p{4cm}lcccccc}
\hline
\textbf{Model} & \textbf{Method} & \textbf{Dimensions} & \textbf{ViDoRe N@5} & \textbf{ViDoRe R@10} & \textbf{Cross N@5} & \textbf{Cross R@10} \\
\hline
Gemma3-LinearMerge (2056) & Linear & 2056 & 54.06 & 63.43 & 73.46 & 89.27 \\
Gemma3-LinearMerge (1536) & Linear & 1536 & 52.04 & 62.01 & 74.77 & 88.53 \\
Gemma3-LinearMerge (768) & Linear & 768 & 48.29 & 58.33 & 70.84 & 80.00 \\
Gemma3-SLERP (2056) & SLERP & 2056 & 53.85 & 63.23 & 74.91 & 89.27 \\
Gemma3-SLERP (1536) & SLERP & 1536 & 51.03 & 61.93 & 74.40 & 88.53 \\
Gemma3-SLERP (768) & SLERP & 768 & 47.42 & 58.38 & 70.86 & 78.53 \\
\hline
\end{tabular}%
}
\end{table}

\begin{figure}[!htbp]
\centering
\includegraphics[width=0.8\textwidth]{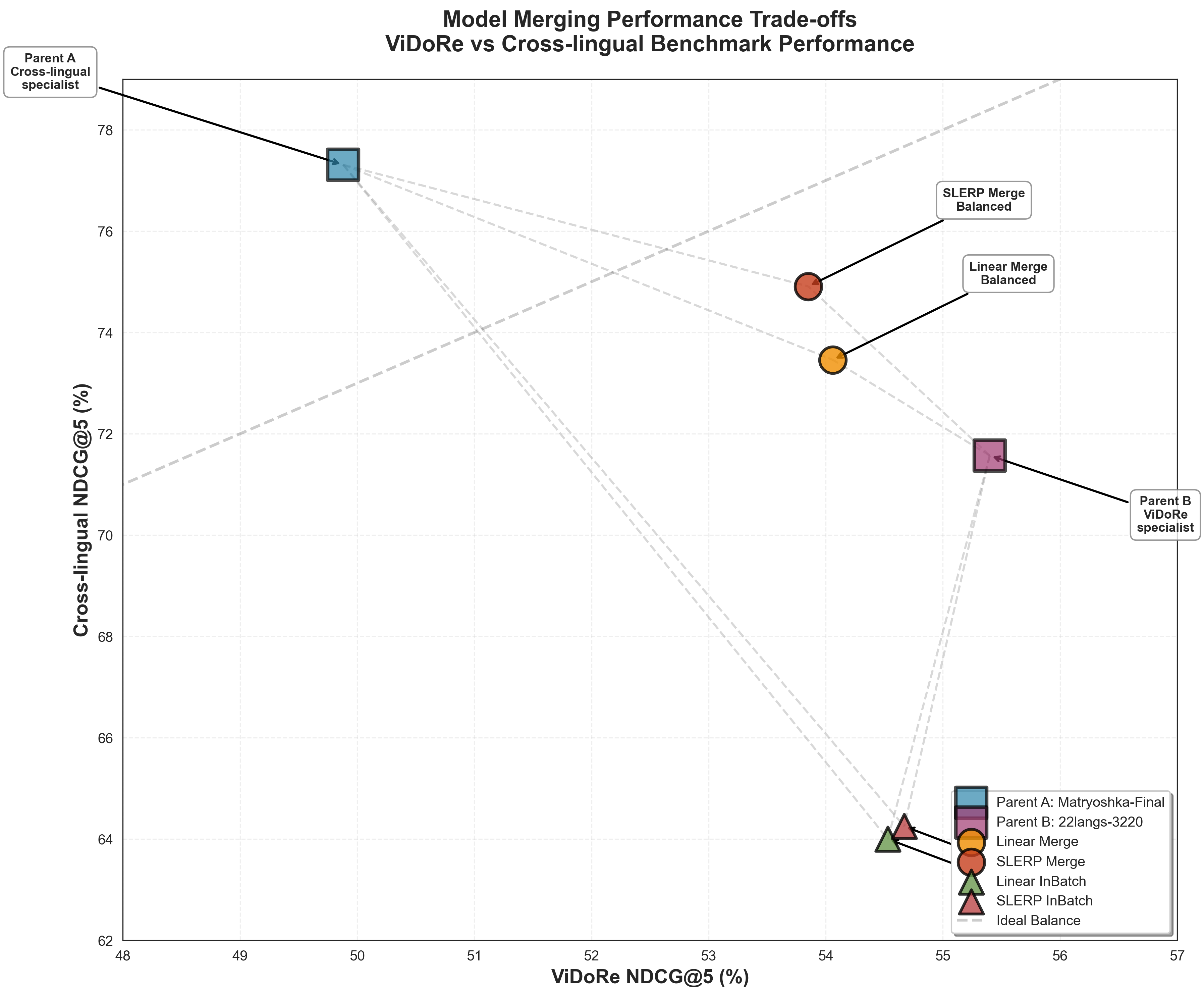}
\caption{\textbf{Model Merging Performance Trade-offs.} Parent models at opposite corners, merged models achieving balanced intermediate performance.}
\label{fig:model_merging_scatter}
\end{figure}

Figure~\ref{fig:model_merging_scatter} visualizes the performance trade-offs through a scatter plot of ViDoRe NDCG@5 versus Cross-lingual NDCG@5. The parent models occupy opposite corners of the performance space: Parent A (Gemma3-CrossLingual) in the upper-left with high cross-lingual but moderate ViDoRe performance, and Parent B (Gemma3-ViDoRe) in the lower-right with high ViDoRe but moderate cross-lingual performance. The merged models successfully populate the balanced middle ground, with Gemma3-SLERP achieving 53.85\% ViDoRe and 74.91\% cross-lingual, and Gemma3-LinearMerge reaching 54.06\% ViDoRe and 73.46\% cross-lingual. The visualization demonstrates that model merging successfully balances objectives without additional training: merged models achieve 95-98\% of the best parent's performance on each benchmark, with SLERP better preserving cross-lingual capabilities and Linear better retaining ViDoRe accuracy.

\FloatBarrier
\section{Scaling to 22 Languages}
\label{appendix:scaling}

\subsection{Dense Models at Full Scale}

\begin{table}[!htbp]
\centering
\caption{\textbf{Dense Model Scaling from 6 to 22 Languages.} SV = Single-Vector.}
\label{tab:dense_scaling_full}
\begin{tabular}{lcccccc}
\hline
\textbf{Model} & \textbf{Type} & \textbf{ViDoRe N@5} & \textbf{ViDoRe R@10} & \textbf{Cross N@5} & \textbf{Cross R@10} \\
\hline
Gemma3-22langs & SV & 55.40 & 63.75 & 71.57 & 87.06 \\
Gemma3-6langs & SV & 49.08 & 59.76 & 60.39 & 72.94 \\
\hline
\end{tabular}%
\end{table}

\begin{figure}[!htbp]
\centering
\includegraphics[width=0.6\textwidth]{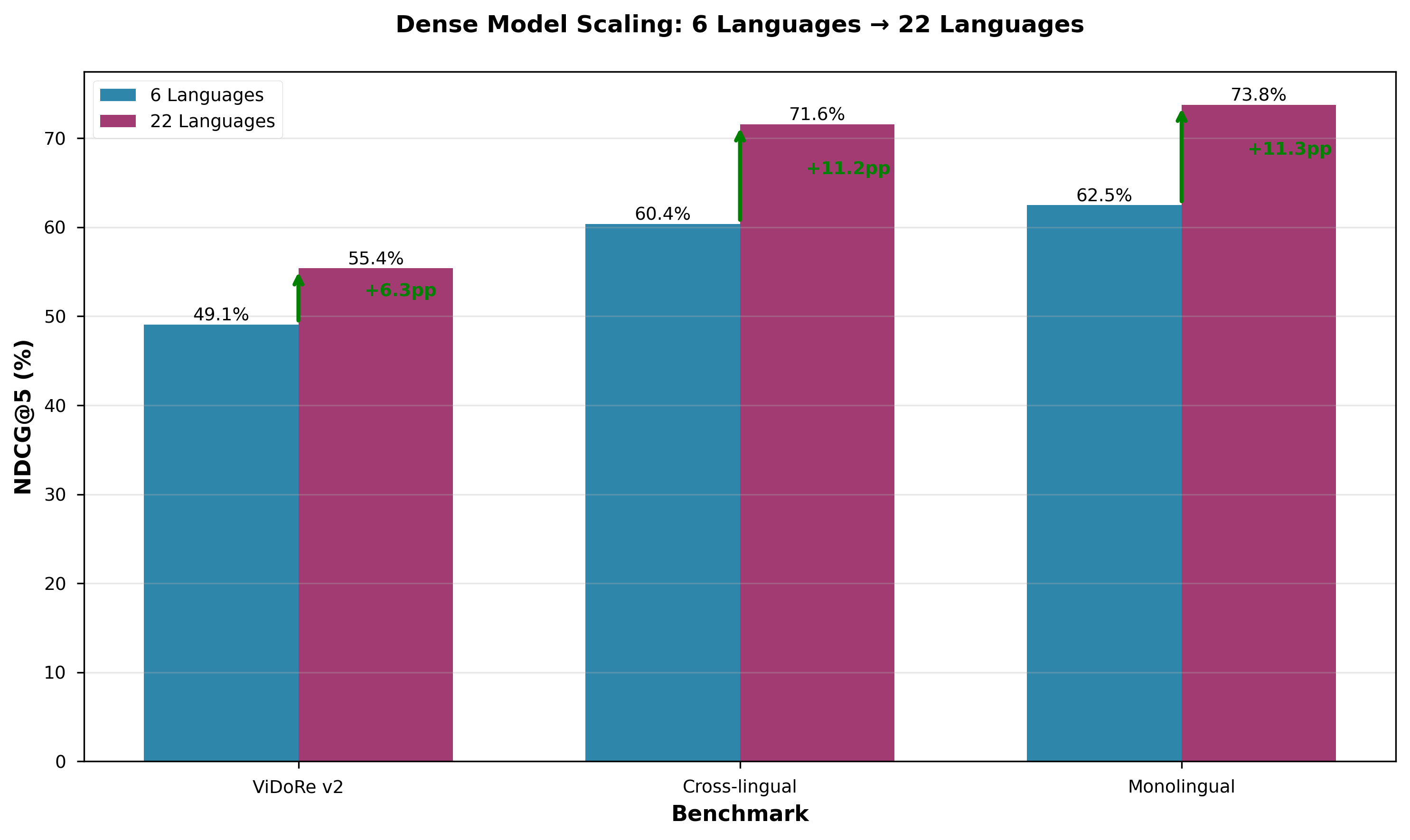}
\caption{\textbf{Language Scaling: Performance improvements when scaling from 6 to 22 languages.}}
\label{fig:scaling_22lang}
\end{figure}

Figure~\ref{fig:scaling_22lang} demonstrates the benefits of linguistic diversity and training data scale. Scaling from 6 to 22 languages yielded substantial improvements: +6.3 points on ViDoRe (55.40\% vs 49.08\%), +11.2 points on cross-lingual tasks (71.57\% vs 60.39\%), and +11.3 points on monolingual tasks. The larger relative gains on cross-lingual and monolingual benchmarks (18\%) versus ViDoRe (13\%) suggest that linguistic diversity particularly benefits non-English retrieval, validating the scaling hypothesis.

\FloatBarrier
\subsection{Multi-Vector Col Models at Full Scale}

\begin{table}[!htbp]
\centering
\caption{\textbf{Col Model Comparison at 22 Languages.} MV = Multi-Vector.}
\label{tab:col_22lang_full}
\resizebox{\textwidth}{!}{%
\begin{tabular}{p{4.5cm}ccccccc}
\hline
\textbf{Model} & \textbf{Type} & \textbf{ViDoRe N@5} & \textbf{ViDoRe R@10} & \textbf{Cross N@5} & \textbf{Cross R@10} & \textbf{Mono N@5} \\
\hline
ColGemma3-Merged-22langs & MV & 55.31 & 67.29 & 63.75 & 69.95 & 67.02 \\
ColGemma3-22langs & MV & 52.91 & 62.13 & 64.74 & 72.35 & 71.21 \\
ColGemma3-ColPaliOnly & MV & 54.37 & 63.60 & 14.15 & 14.85 & N/A \\
ColQwen2-22langs & MV & 53.07 & 62.72 & 42.86 & 52.01 & 59.08 \\
ColQwen2-22langs-Scaled & MV & N/A & N/A & 51.08 & 60.45 & 64.31 \\
ColPali-22langs & MV & 50.37 & 61.83 & 42.64 & 53.63 & 53.38 \\
\hline
\multicolumn{7}{l}{\textit{For comparison (6-language baselines):}} \\
ColQwen2-6langs & MV & 58.63 & 67.62 & 37.15 & 39.85 & N/A \\
ColPali-6langs & MV & 56.41 & 64.94 & 42.22 & 50.05 & 53.38 \\
\hline
\end{tabular}%
}
\end{table}

Multi-vector ColGemma3 trained from scratch showed significantly better cross-lingual generalization (63.75\% NDCG@5) compared to finetuned ColPali (42.64\%) and ColQwen2 (42.86\%). The final model followed a three-stage process: Stage 1 trained on ColPali data (54.37\% ViDoRe, 14.15\% cross-lingual), Stage 2 finetuned on Nayana IR 22 languages (52.91\% ViDoRe, 64.74\% cross-lingual), and Stage 3 merged the checkpoints (55.31\% ViDoRe, 63.75\% cross-lingual), creating the most balanced late interaction model. Interestingly, the 6-language ColQwen2 model outperformed the 22-language version on ViDoRe (58.63\% vs 53.07\%), suggesting potential overfitting or training instability.

\FloatBarrier
\section{Dense vs Col Architecture Comparison}
\label{appendix:dense_vs_col}

We trained models using both retrieval architectures to evaluate trade-offs: single-vector models enabling fast vector similarity search, and multi-vector late interaction models using multiple embeddings with MaxSim matching.

\begin{table}[!htbp]
\centering
\caption{\textbf{Dense vs. Late Interaction Architecture Performance.} SV = Single-Vector; MV = Multi-Vector.}
\label{tab:dense_vs_col_performance}
\resizebox{\textwidth}{!}{%
\begin{tabular}{p{2cm}p{4.5cm}cccccc}
\hline
\textbf{Type} & \textbf{Model} & \textbf{ViDoRe N@5} & \textbf{ViDoRe R@10} & \textbf{Cross N@5} & \textbf{Cross R@10} & \textbf{Mono N@5} \\
\hline
SV & Gemma3-LinearMerge & 54.06 & 63.43 & 73.46 & 89.27 & N/A \\
SV & Gemma3-SLERP & 53.85 & 63.23 & 74.91 & 89.27 & N/A \\
SV & Gemma3-Matryoshka & 49.88 & 62.00 & 77.31 & 88.38 & 74.10 \\
MV & ColGemma3-Merged & 55.31 & 67.29 & 63.75 & 69.95 & 67.02 \\
MV & ColGemma3-22langs & 52.91 & 62.13 & 64.74 & 72.35 & 71.21 \\
\hline
\end{tabular}%
}
\end{table}

\begin{table}[!htbp]
\centering
\caption{\textbf{Dense vs. Late Interaction: Deployment Trade-offs.}}
\label{tab:dense_vs_col_tradeoffs_full}
\begin{tabular}{p{3.5cm}p{4.5cm}p{4.5cm}}
\hline
\textbf{Aspect} & \textbf{Dense (Single-Vector)} & \textbf{Col (Multi-Vector)} \\
\hline
Storage & 2056 dims $\approx$ 8KB & 128 tokens $\times$ 128 dims $\approx$ 64KB \\
Storage advantage & \textbf{8-10$\times$ smaller} & - \\
Retrieval speed & 500-1000 QPS & 50-100 QPS \\
Speed advantage & \textbf{10$\times$ faster} & - \\
Cross-lingual perf & \textbf{73-77\% NDCG@5} & 64\% NDCG@5 \\
ViDoRe perf & 50-54\% NDCG@5 & \textbf{53-55\% (marginal)} \\
Interpretability & Low (black-box vector) & \textbf{High (attention maps)} \\
Long-tail queries & Moderate & \textbf{Better (multi-vector)} \\
Deployment & \textbf{Simple (standard database)} & Complex (MaxSim) \\
\hline
\end{tabular}%
\end{table}

The architectural comparison reveals complementary strengths. Single-vector models achieved 73-77\% NDCG@5 on cross-lingual tasks versus 64\% for late interaction, a 10-13 point advantage. Late interaction models showed marginal superiority on ViDoRe (55.31\% vs 53-55\%), a 0-2 point difference. Beyond retrieval accuracy, single-vector embeddings dominate on practical deployment metrics: 8-10$\times$ smaller storage footprint enabling larger indices and lower costs; 10$\times$ faster retrieval (standard vector similarity vs expensive MaxSim across 128 token vectors); and compatibility with standard vector databases (Faiss, Milvus, Pinecone) simplifying infrastructure. Late interaction's advantages lie in interpretability through token-level attention heatmaps and potentially better handling of long-tail queries through fine-grained multi-vector matching. Both architectures have merit depending on deployment priorities: single-vector for cost-efficiency and scale, late interaction for interpretability and explainability.

\FloatBarrier
\section{Cross-Lingual Embedding Convergence: PCA Analysis}
\label{appendix:convergence}

This section provides visual evidence that the Gemma3 model progressively learns cross-lingual alignment during training on the Nayana IR dataset. We visualize how embeddings for semantically equivalent content across different languages evolve from language-separated clusters to semantically-aligned representations.

\FloatBarrier
\subsection{Methodology}

We extract model checkpoints at regular intervals (500, 1500, 2500, 3500, 4500, 5066 steps), generate embeddings for document images paired with queries in multiple languages, apply PCA to project 2560-dimensional embeddings to 2D, and evaluate convergence behavior across different multilingual configurations (2, 6, and 15 languages).

\FloatBarrier
\subsection{2-Language Alignment: Kannada $\leftrightarrow$ English}

\begin{figure}[!htbp]
\centering
\includegraphics[width=\textwidth]{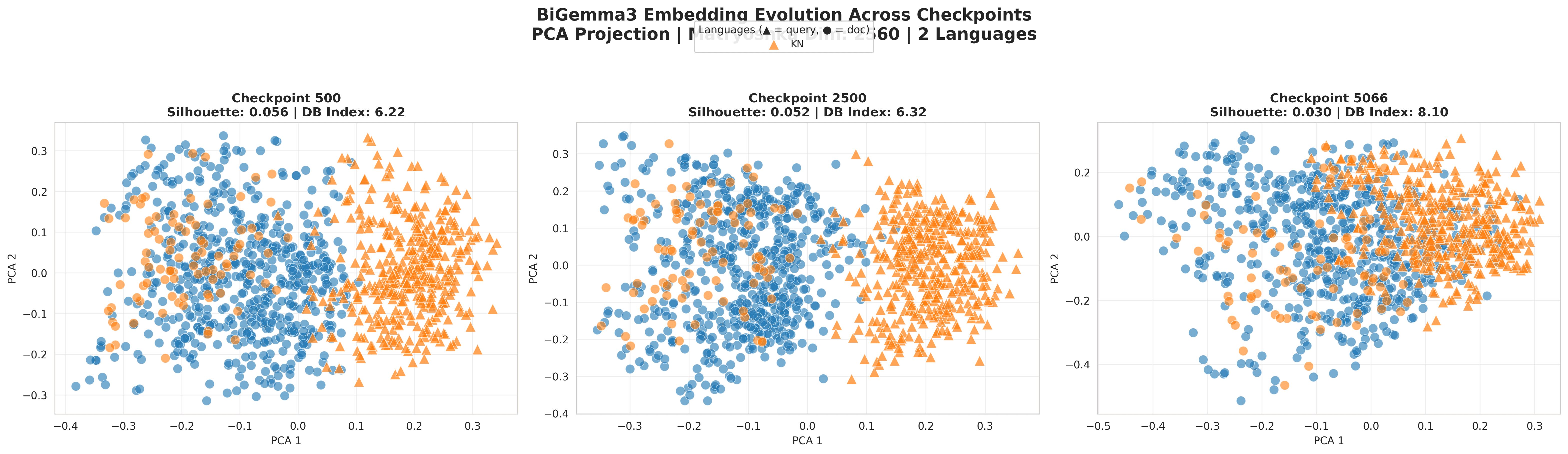}
\caption{\textbf{Gemma3 Embedding Evolution: 2 Languages (KN Query $\leftrightarrow$ Doc).}}
\label{fig:convergence_2lang}
\end{figure}

Figure~\ref{fig:convergence_2lang} visualizes the simplest case of cross-lingual alignment between Kannada queries and English documents across six training checkpoints. At checkpoint 500 (leftmost panel), the PCA projection shows clear separation between blue triangles (Kannada queries) and orange triangles (document embeddings), forming distinct clusters in opposite regions of the embedding space. As training progresses to checkpoint 2500 (middle panels), the visualization reveals boundaries beginning to blur as query and document embeddings start overlapping in the center of the plot. By checkpoint 5066 (rightmost panel), the separation has largely dissolved: blue and orange triangles intermingle throughout the embedding space, indicating the model learned to produce similar embeddings for semantically equivalent content regardless of language. This progressive convergence from language-separated to semantically-aligned representations demonstrates the model's acquisition of cross-lingual understanding.

\FloatBarrier
\subsection{6-Language Cross-Lingual Alignment}

\begin{figure}[!htbp]
\centering
\includegraphics[width=\textwidth]{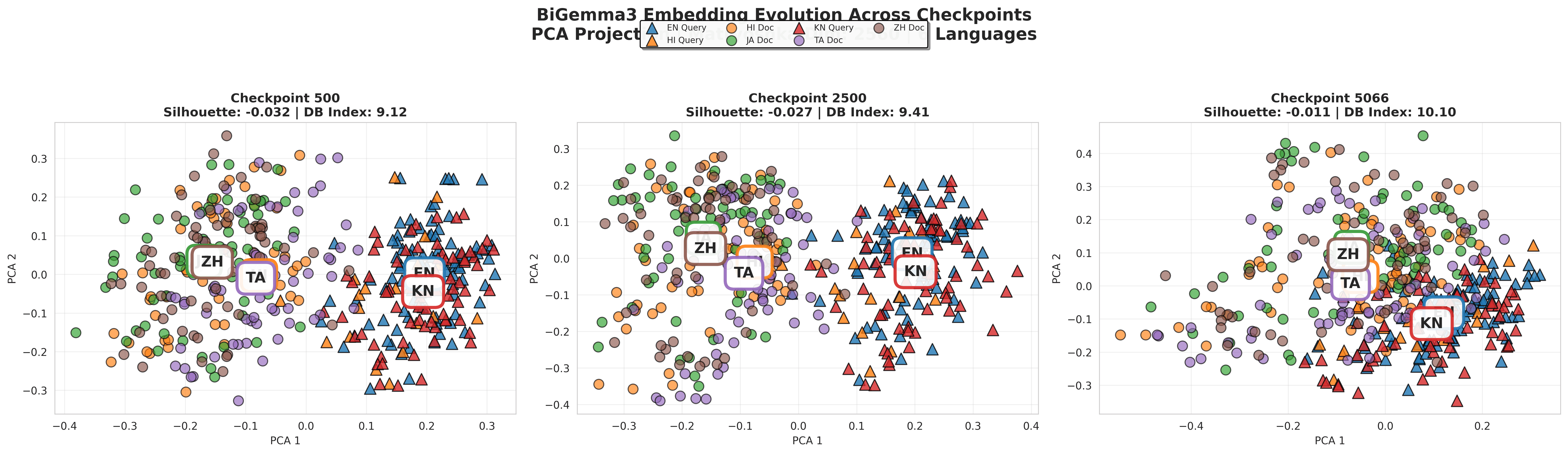}
\caption{\textbf{Gemma3 Embedding Evolution: 6 Languages.}}
\label{fig:convergence_6lang}
\end{figure}

Scaling to six languages (English, Hindi, Chinese, Tamil, Japanese, Kannada) reveals richer dynamics of cross-lingual convergence. Figure~\ref{fig:convergence_6lang} shows six distinct color-coded language clusters clearly visible at checkpoint 500, with each language occupying its own region in the 2D PCA space. Queries (triangles) and documents (circles) form separate but overlapping zones. As training advances through checkpoints 1500 and 2500, the language-specific boundaries progressively dissolve i.e., the distinct color clusters merge toward the center of the plot. By checkpoint 5066, the visualization displays near-complete cross-lingual mixing: embeddings from different languages for the same semantic content cluster together regardless of script, with Hindi (Devanagari), Chinese (Hanzi), Tamil, Kannada, Japanese (Kanji/Kana), and English (Latin) all converging to similar regions. The transition from distinct language-separated clusters to a unified semantic space provides direct visual evidence that the model learns to align representations across diverse writing systems.

\begin{figure}[!htbp]
\centering
\includegraphics[width=\textwidth]{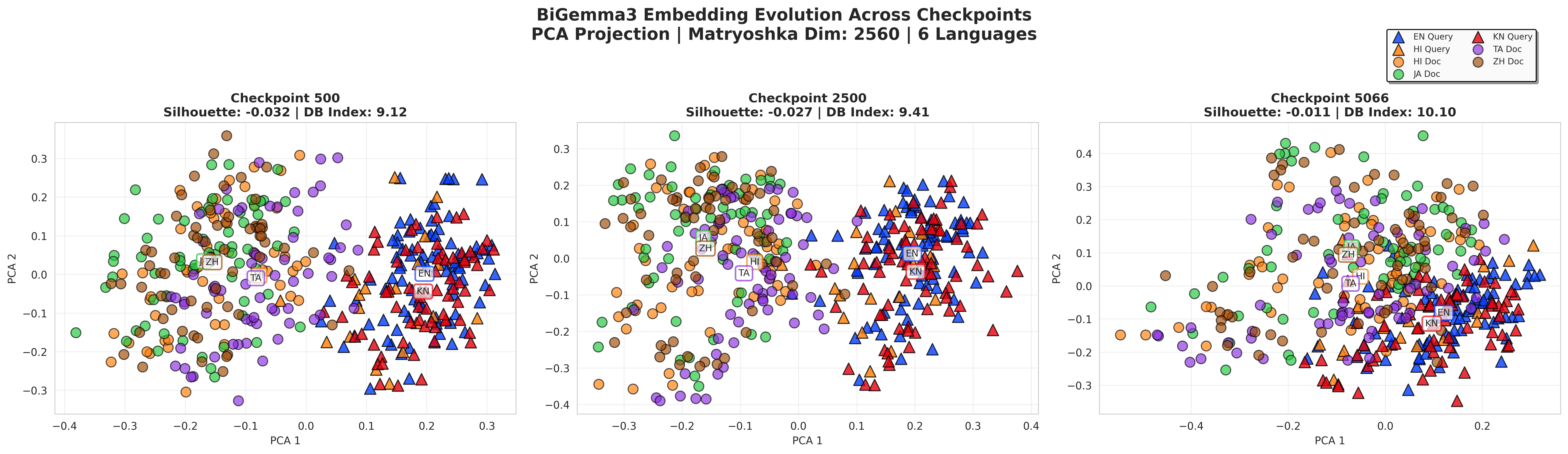}
\caption{\textbf{Gemma3 Embedding Evolution: 6 Languages (Matryoshka Dim: 2560).}}
\label{fig:convergence_6lang_matryoshka}
\end{figure}

Figure~\ref{fig:convergence_6lang_matryoshka} confirms the same convergence pattern using Matryoshka embeddings with 2560 dimensions. The visualization mirrors the behavior observed in the standard model: early checkpoints show language-clustered distributions, which progressively merge into semantically-organized representations. This demonstrates that Matryoshka's multi-scale training objective preserves (and potentially enhances) cross-lingual alignment quality, validating that flexible dimensionality does not compromise the model's ability to learn language-agnostic representations.

\subsection{15-Language Scaling}

\begin{figure}[!htbp]
\centering
\includegraphics[width=\textwidth]{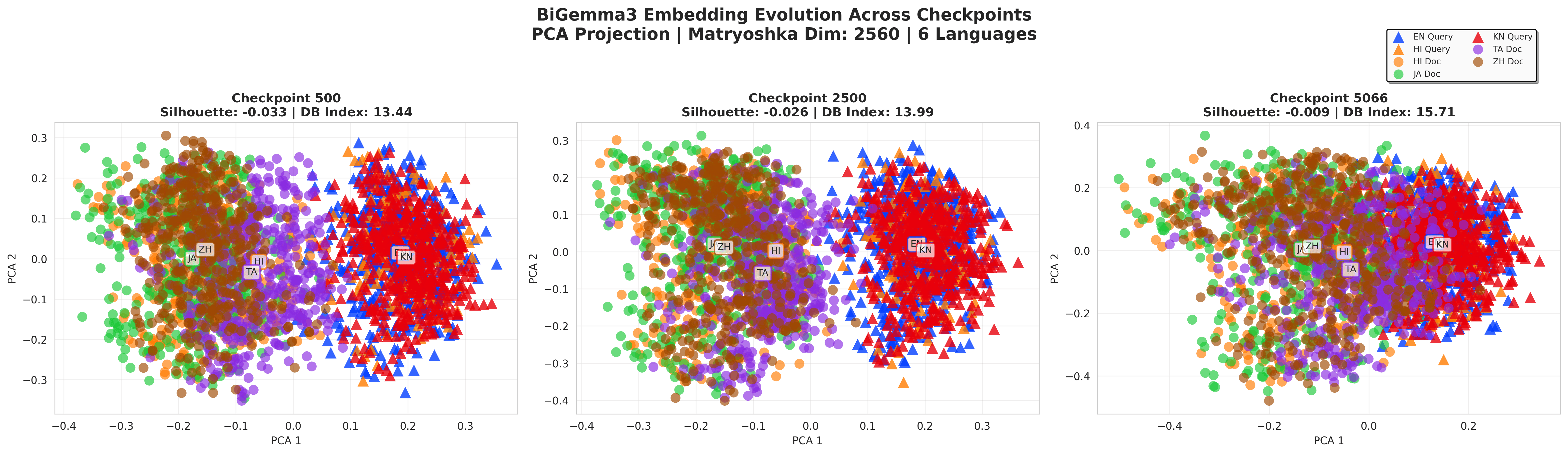}
\caption{\textbf{Gemma3 Embedding Evolution: 15 Languages.}}
\label{fig:convergence_15lang}
\end{figure}

Figure~\ref{fig:convergence_15lang} tests the limits of cross-lingual alignment with 15 diverse languages spanning multiple language families and writing systems: Arabic, Bengali, German, Spanish, French, Hindi, Kannada, Russian, Tamil, Telugu, Thai, and Chinese. At checkpoint 500, the PCA projection shows significant dispersion across the entire embedding space, with language families forming loose clusters (e.g., Indic languages grouping together, Romance languages clustering separately). The visualization tracks progressive convergence: checkpoint 1500 shows early signals of alignment with document embeddings beginning to form denser central clusters; checkpoint 2500 reveals clear movement toward the center with query embeddings from different languages increasingly overlapping; and checkpoint 5066 demonstrates strong cross-lingual mixing across all 15 languages despite the complexity. The model successfully learns to align semantically equivalent content even when confronted with high linguistic diversity, demonstrating the robustness of the M3DR framework to scale beyond language pairs to truly multilingual scenarios.

\FloatBarrier
\subsection{Hindi $\leftrightarrow$ Kannada Detailed Analysis}

\begin{figure}[!htbp]
\centering
\includegraphics[width=\textwidth]{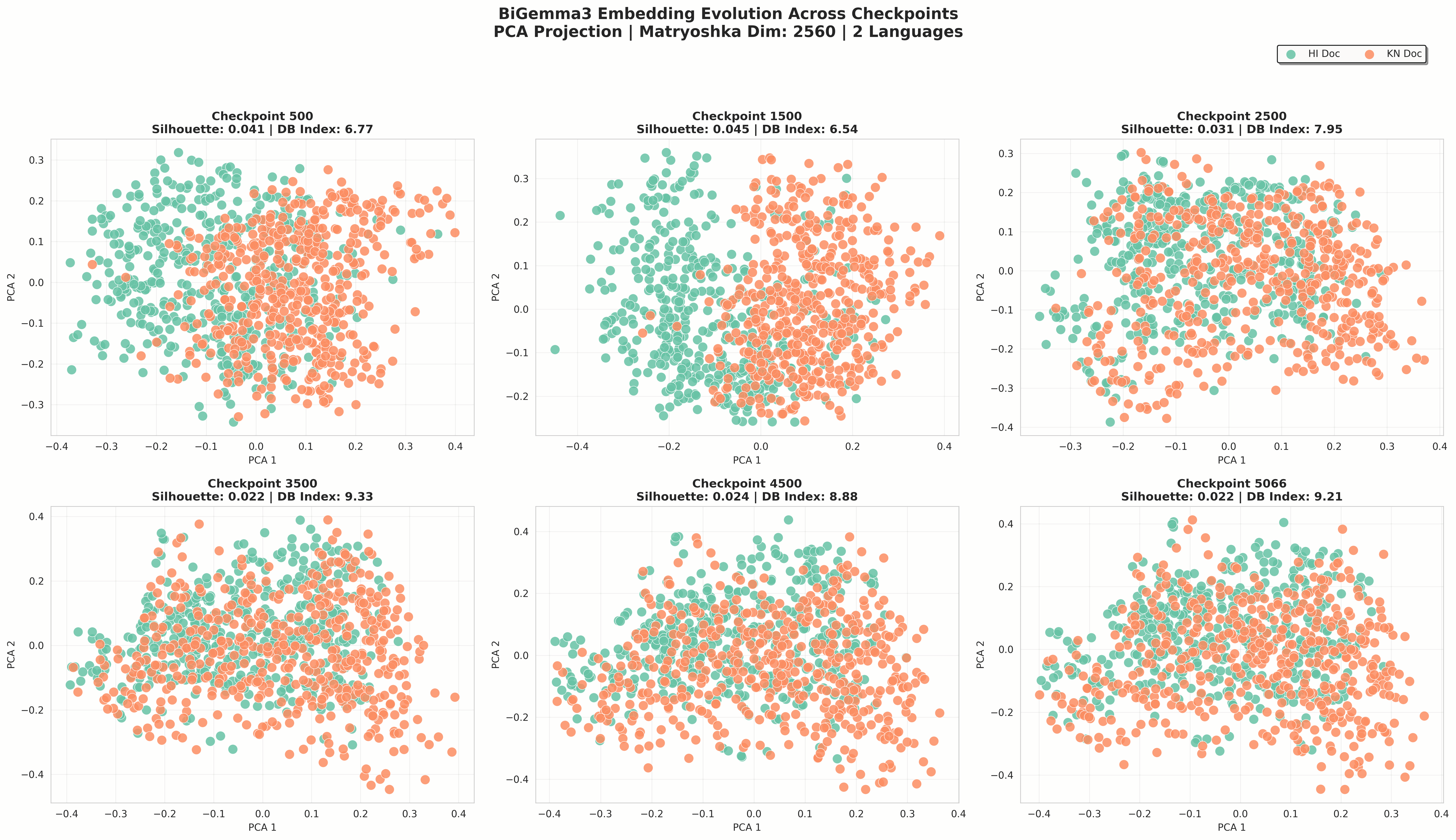}
\caption{\textbf{Gemma3 Embedding Convergence: Hindi $\leftrightarrow$ Kannada.}}
\label{fig:convergence_hi_kn}
\end{figure}

Figure~\ref{fig:convergence_hi_kn} provides focused analysis of two Indic languages with distinct scripts to demonstrate fine-grained cross-lingual alignment. At checkpoint 500, Hindi documents (green circles) and Kannada documents (orange circles) are clearly separated in the PCA space, occupying opposite regions of the plot. The progression through checkpoints 1500 and 2500 shows gradual overlap beginning between the clusters, with the color boundaries becoming less distinct. By checkpoint 5066, the visualization reveals near-perfect alignment: green and orange circles are thoroughly intermixed throughout the embedding space, demonstrating that the model learned to produce nearly identical embeddings for the same visual content whether queried in Hindi (Devanagari script) or Kannada (Kannada script). This alignment between non-Latin scripts provides compelling evidence that the model develops script-agnostic semantic representations rather than superficial pattern matching.

\subsection{Implications for Cross-Lingual Retrieval}

These visualizations collectively demonstrate that multilingual training with the Nayana IR dataset enables models to learn progressive cross-lingual alignment in the embedding space. The convergence happens gradually across training rather than as a sudden phase transition, with the model learning to project different languages into a shared semantic space. Despite using single-vector representations, these embeddings successfully capture cross-lingual semantics, as evidenced by the visual collapse of language-specific clustering into semantically-organized distributions. This validates our core hypothesis: the Nayana IR dataset contains sufficient cross-lingual signal to enable robust multilingual multimodal retrieval across diverse scripts and language families.

\section{Col Model Attention Visualization}
\label{appendix:col_attention}

Col models provide interpretability through token-level attention heatmaps via maximum similarity (MaxSim) maps between query tokens and image patches. We demonstrate that cross-lingual text queries attend to the same visual regions, validating language-agnostic visual grounding.

\FloatBarrier
\subsection{Experimental Setup}

We compare models trained solely on English-centric data (ColPali base versions) against those fine-tuned on the multilingual Nayana corpus. The test document contains the scientific term \textit{``Acremonium coephenophialum''} in English, queried in three languages - English, Hindi and Kannada.

\begin{figure}[!htbp]
\centering
\includegraphics[width=\textwidth]{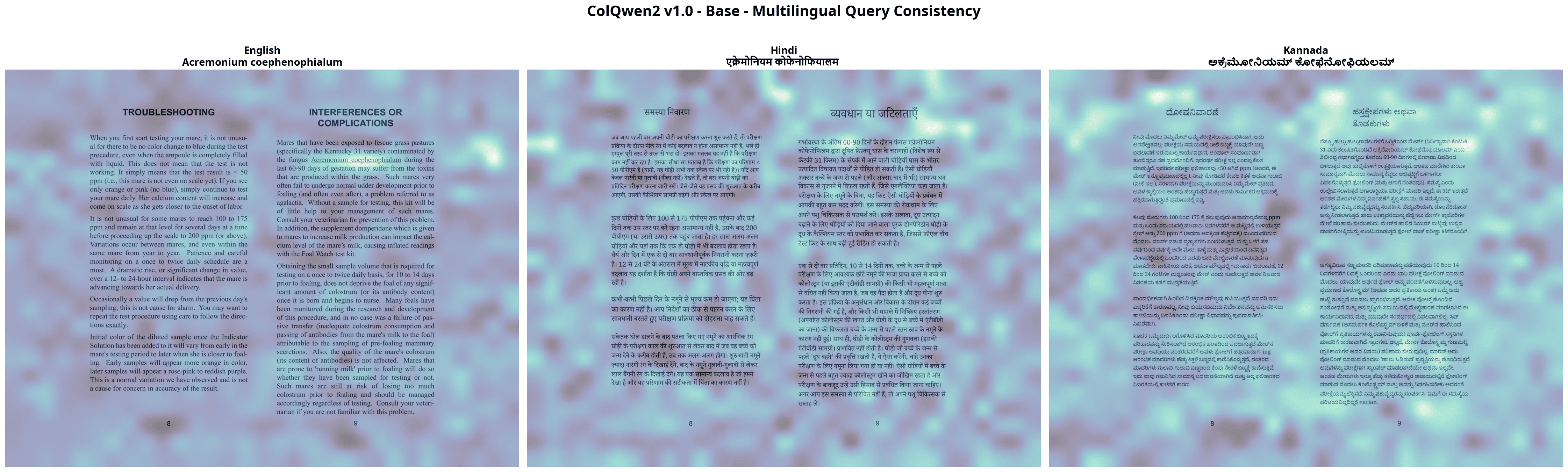}
\caption{\textbf{ColQwen2-Base (MV).} English Max Sim: 0.335 | Hindi: 0.247 | Kannada: 0.286}
\label{fig:colqwen_base}
\end{figure}

\begin{figure}[!htbp]
\centering
\includegraphics[width=\textwidth]{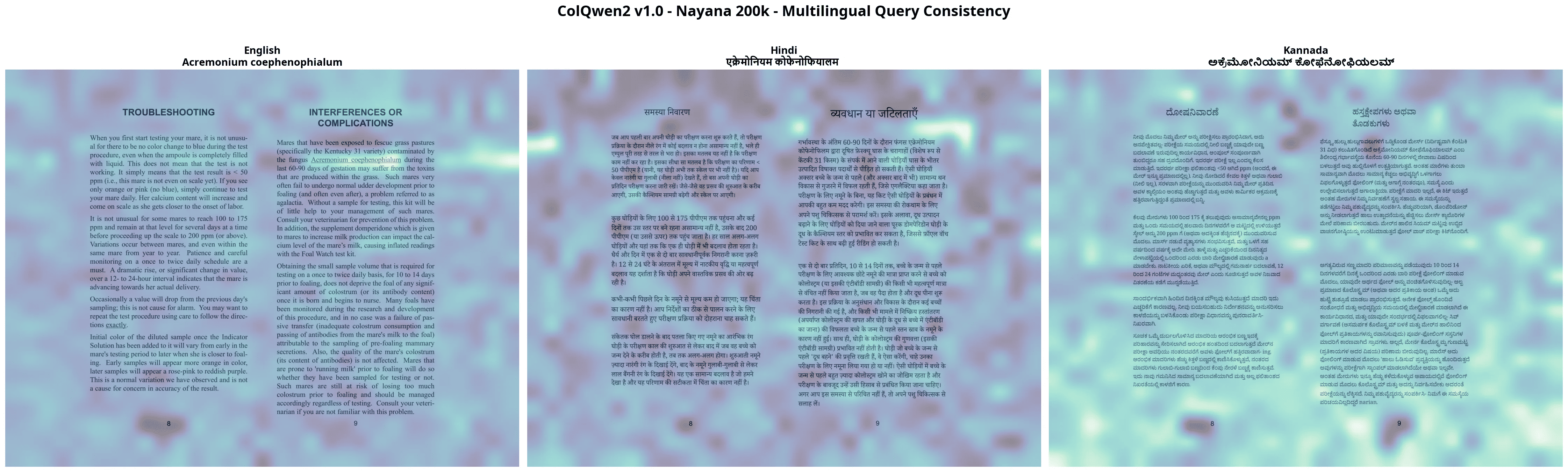}
\caption{\textbf{ColQwen2-Finetuned (MV) - Ours.} English Max Sim: 0.529 | Hindi: 0.436 | Kannada: 0.269}
\label{fig:colqwen_nayana}
\end{figure}

\begin{figure}[!htbp]
\centering
\includegraphics[width=\textwidth]{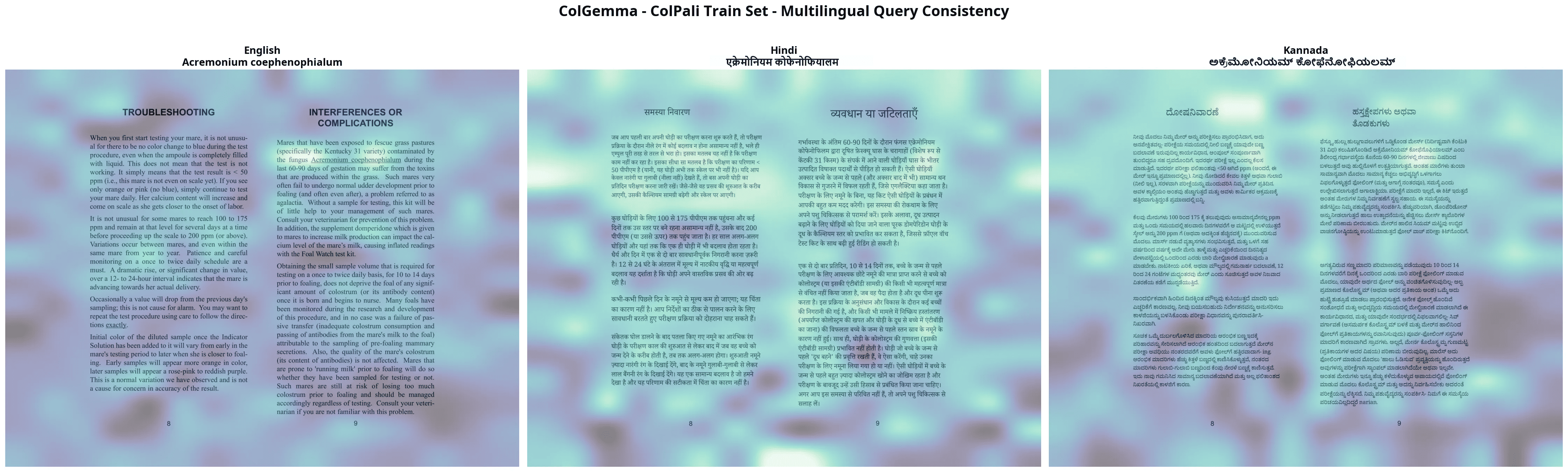}
\caption{\textbf{ColGemma3-ColPaliOnly (MV).} English Max Sim: 0.329 | Hindi: 0.203 | Kannada: 0.109}
\label{fig:colgemma_base}
\end{figure}

\begin{figure}[!htbp]
\centering
\includegraphics[width=\textwidth]{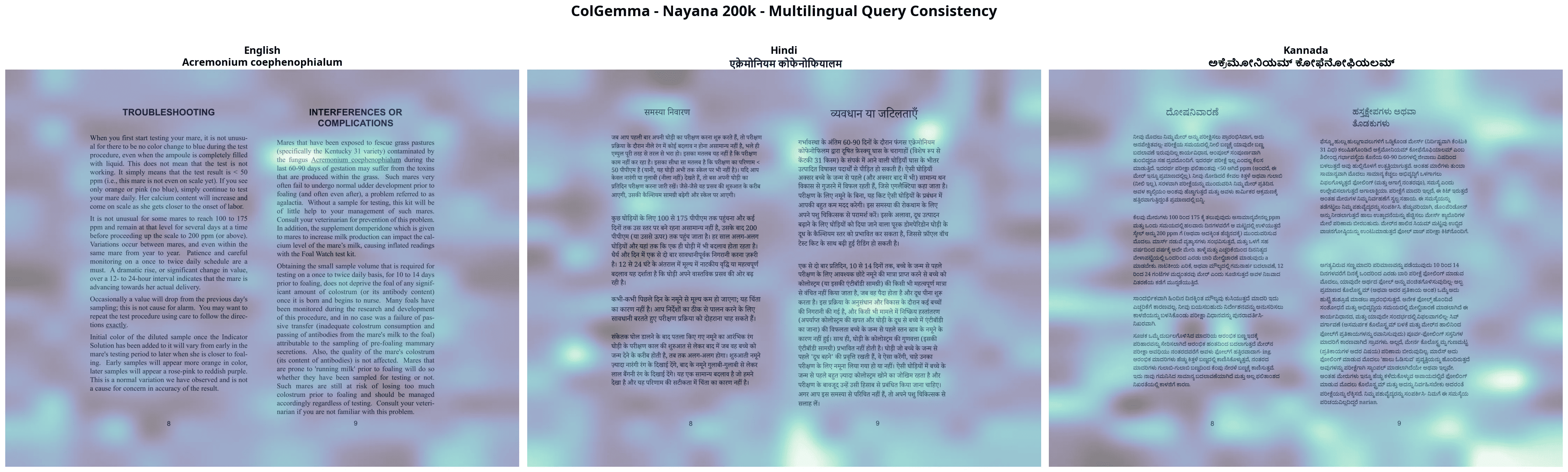}
\caption{\textbf{ColGemma3-Finetuned (MV) - Ours.} \\ English Max Sim: 0.335 | Hindi: 0.381 | Kannada: 0.349}
\label{fig:colgemma_nayana}
\end{figure}

Figures~\ref{fig:colqwen_base} through~\ref{fig:colgemma_nayana} reveal the impact of multilingual fine-tuning on attention consistency. The base models (Figures~\ref{fig:colqwen_base} and~\ref{fig:colgemma_base}) show attention heatmaps where English queries correctly focus on the target text with moderate confidence (red intensity), but Hindi and Kannada queries either diffuse attention across the document or fail to activate the correct region, ColGemma3-ColPaliOnly drops from 0.329 similarity in English to just 0.109 in Kannada, indicating near-complete failure to associate the Kannada script with the English visual text. In contrast, the Nayana-tuned models (Figures~\ref{fig:colqwen_nayana} and~\ref{fig:colgemma_nayana}) demonstrate query consistency: all three language queries activate precisely the same visual regions with comparable confidence levels (red heatmap intensity), proving the models learned robust cross-lingual representations. Remarkably, ColGemma3-Finetuned achieves higher Hindi similarity (0.381) than English (0.335), demonstrating script-agnostic semantic understanding rather than relying on visual script matching.

\begin{table}[!htbp]
\centering
\caption{\textbf{Cross-Lingual Query Consistency Analysis.} MV = Multi-Vector.}
\label{tab:col_attention_consistency}
\begin{tabular}{lcccc}
\hline
\textbf{Model} & \textbf{English MaxSim} & \textbf{Hindi MaxSim} & \textbf{Kannada MaxSim} & \textbf{Cross-Lingual Gap} \\
\hline
ColGemma3-ColPaliOnly & 0.329 & 0.203 & 0.109 & \textcolor{red}{-0.220} \\
ColGemma3-Finetuned & 0.335 & \textbf{0.381} & \textbf{0.349} & \textcolor{green}{+0.015} \\
ColQwen2-Base & 0.335 & 0.247 & 0.286 & -0.088 \\
ColQwen2-Finetuned & \textbf{0.529} & \textbf{0.436} & 0.269 & -0.260 \\
\hline
\end{tabular}%
\end{table}

Table~\ref{tab:col_attention_consistency} quantifies the dramatic improvement in cross-lingual consistency. ColGemma3-Finetuned achieves remarkable query consistency with only a +0.015 gap between English and the lowest language score, even showing higher Hindi scores (0.381) than English (0.329). This contrasts starkly with ColGemma3-ColPaliOnly which exhibits catastrophic cross-lingual failure with a -0.220 gap and Kannada performance plummeting to 0.109. The Nayana-tuned models demonstrate that multilingual fine-tuning does not degrade English capabilities while dramatically improving non-English query performance, with all language queries activating identical visual regions which is clear evidence of true cross-lingual semantic understanding rather than script-matching heuristics.

\FloatBarrier
\section{Training Configuration and Computational Cost}
\label{appendix:training_config}

\subsection{LoRA Setup}
Models were finetuned with LoRA (rank 32, alpha 32, dropout 0.1) applied to projection layers (down\_proj, gate\_proj, up\_proj, k\_proj, q\_proj, v\_proj, o\_proj).

\subsection{Hyperparameters}
Training for 2 epochs gave the best results. Settings: per device batch size 32, learning rate 2e-4, gradient checkpointing, warmup 100 steps, logging 10 steps, and saving every 500 steps.

\subsection{Compute}
Small runs used one A100 80 GB for 2 to 4 hours on roughly 45k pairs. Final multilingual training used 4 to 8 A100 80 GB GPUs with DDP and mixed precision. Two datasets were trained for 2 epochs each which took 6 to 8 hours in total (about 64 GPU hours, roughly 100 to 150 USD).

\subsection{Data}
\begin{table}[!htbp]
\centering
\caption{\textbf{Dataset Summary}}
\label{tab:dataset_statistics}
\begin{tabular}{lcc}
\hline
\textbf{Dataset} & \textbf{Pairs} & \textbf{Notes} \\
\hline
ColPali & \(\sim\)150k & English document retrieval \\
Nayana IR & \(\sim\)250k & 22 languages, cross lingual \\
\textbf{Total} & \textbf{\(\sim\)400k} & Multilingual retrieval \\
\hline
\end{tabular}
\end{table}

Nayana IR spans Indic scripts (11), major European languages (6), Asian languages (4), and Arabic. This coverage supports both multilingual and cross lingual document retrieval tasks.

\FloatBarrier


\end{document}